\documentclass[
 manuscript%,
]{aastex63}
\usepackage{amsmath}
\usepackage{float}
\usepackage{multirow}

\received{}
\revised{}
\accepted{}
\shorttitle{Skew-Kappa electron heat-flux}
\shortauthors{Zenteno-Quinteros, Vi\~nas \& Moya}

\begin{document}
	
	\title{Skew-Kappa distribution functions \& whistler-heat-flux instability in the solar wind: the core-strahlo model}
	
	\author[0000-0003-2430-6058]{Bea Zenteno-Quinteros}
	\email{beatriz.zenteno@ug.uchile.cl}
	\affiliation{Departamento de F\'isica, Facultad de Ciencias, Universidad de Chile, Santiago, Chile}
	
	\author[0000-0001-5912-5703]{Adolfo F. Vi\~nas}
	\affiliation{Department of Physics \& the Institute for Astrophysics and Computational Sciences (IACS), Catholic University of America, Washington-DC, 20064, USA}  
	\affiliation{NASA Goddard Space Flight Center, Heliospheric Science Division, Geospace Physics Laboratory, Mail Code 673, Greenbelt, MD 20771, USA}
	\author[0000-0002-9161-0888]{Pablo S. Moya}
	\email{pablo.moya@uchile.cl}
	\affiliation{Departamento de F\'isica, Facultad de Ciencias, Universidad de Chile, Santiago, Chile}
	
	
	\begin{abstract}
        Electron velocity distributions in the solar wind are known to have field-aligned skewness, which has been characterized by the presence of secondary populations such as the halo and strahl. Skewness may provide energy for the excitation of electromagnetic instabilities, such as the whistler heat-flux instability (WHFI), that may play an important role in regulating the electron heat-flux in the solar wind. Here we use kinetic theory to analyze the stability of the WHFI in a solar-wind-like plasma where solar wind core, halo and strahl electrons are described as a superposition of two distributions: a Maxwellian core, and another population modeled by a Kappa distribution to which an asymmetry term has been added, representing the halo and also the strahl. Considering distributions with small skewness we solve the dispersion relation for the parallel propagating whistler-mode and study its linear stability for different plasma parameters. Our results show that the WHFI can develop in this system, and provide stability thresholds for this instability, as a function of the electron beta and the parallel electron heat-flux, to be compared with observational data. However, since different plasma states, with different stability level to the WHFI, can have the same moment heat-flux value, it is the skewness (i.e. the asymmetry of the distribution along the magnetic field), and not the heat-flux, the best indicator of instabilities. Thus, systems with high heat-flux can be stable enough to WHFI, so that it is not clear if the instability can effectively regulate the heat-flux values through wave-particle interactions.
	\end{abstract}
	
	%\onecolumn \maketitle %\normalsize \vfill
	
%%%%%%%%%%%%%%%%%%%%%%%%% INTRODUCTION %%%%%%%%%%%%%%%%%%

\section{Introduction}
\label{sec:introduction}
	
Space plasmas are magnetized systems that can be out of thermal equilibrium due to the low collision frequency between its constituent particles. Coulomb collisions are an efficient mechanism to relax particle populations to thermodynamic equilibrium, where the distribution functions reduce to Maxwellian profiles. Therefore, on collisionless systems the particles velocity distribution function can develop non-thermal features. These non-thermal characteristics represent free energy in the system, that can be emitted as electromagnetic radiation, such that the plasma relaxes to more stable states via non-collisional processes as wave particle interactions.
		
The solar wind shows several non-thermal features in the electron velocity distribution (eVDF). Among them, the field-aligned skewness \citep{feldman1975solar,ScudderOlbert79,marsch1982solar,pilipp1987variations,salem2003electron,marsch2004temperature,nieves2008solar} is clearly observed. This asymmetry provides the energy to excite different modes, depending on the plasma parameters \citep{gary1975heat,shaaban2018clarifying}. One of these modes correspond to the so-called  whistler heat-flux instability (WHFI), and the excitation of this branch is due to the free energy provided by the skewness or asymmetry of the eVDF. The WHFI has received attention throughout the years because the associated whistler waves are one of the main candidates for a non-collisional regulation of the electron heat-flux values in the solar wind \citep{abraham1977whistler, gary1994whistler, gary2000whistler,scime1994regulation, scime2001solar,  lacombe2014whistler, kuzichev2019nonlinear, shaaban2019quasi}.
		
There is ample observational evidence that the electron heat-flux in the solar wind cannot be fully described by the collisional Spitzer-Harm theory \citep{spitzer1953transport}. This model is able to adequately describe measurements of the heat-flux under slow solar wind conditions, but in many other cases the Spitzer-Harm law predicts higher values than those shown by in-situ measurements at 1 AU from the Sun \citep{bale2013electron}. This phenomenon of depletion of the electron heat-flux below the values predicted by the collisional transport model has been studied for decades. For example, empirical calculations have been carried out in order to reproduce the measured heat-flux values through an ad-hoc reduction of the thermal conductivity~\citep{cuperman1972characteristics}. Theoretical models have been also proposed, considering different physical mechanisms that can potentially regulate the electron heat-flux through collisionless mechanisms~\citep{forslund1970instabilities,hollweg1972heat,landi2012competition}. The most accepted mechanism to explain this suppression corresponds to a non-collisional regulation due to kinetic process of wave-particle interaction \citep{scime1994regulation, hollweg1974electron,perkins1973heat}. The main candidate for constraining the electron heat-flux values are the whistler waves excited by the heat-flux instability, however, the dominant wave mode involved in this non collisional regulation process is still under debate \citep{gary1977solar, scime2001solar, bale2013electron, shaaban2018beaming, lopez2020alternative}.
		
Most studies of the solar wind electron skewness assumed that the electron populations are composed of different sub-populations; each modelled by a Maxwellian, Bi-Maxwellian or Kappa distribution, which combined can form a skew non-thermal distribution. An example of this in the solar wind is the linear superposition of the drifting electron core, halo and strahl populations \citep{stverak2009radial, saeed2016electron,lazar2018electromagnetic,lopez2020alternative}. Note that these typical distribution functions are symmetrical by themselves, and do not show any skewness. Under this context, in recent years studies have been developed where less common functions are used to model the electron sub-populations. The particularity of these “new” ad-hoc functions is that they are, in fact, asymmetrical. For example, in \citet{horaites2018stability} the authors analyze the kinetic stability of a plasma where the strahl population is described by an analytic function, which was derived from the collisional kinetic equation. Also, in \citet{vasko2019whistler} the authors modeled the strahl population by means of a bi-Maxwellian function to which extra parameters were added that allows the modification of its symmetry. 

Along the same line, here we propose a new heuristic model for solar wind electrons, that can reproduce the behaviour of a core-halo-strahl representation but using only two sub-populations: a bi-Maxwellian core plus a modification to the Kappa distribution that introduces skewness, representing the halo and strahl electrons in a single skew distribution. This skew-Kappa distribution was first proposed by \citet{beck2000application} in a study of fluid turbulence. In the original derivation, the author showed that the asymmetry of the VDF is related with the level of turbulence of the media, measured by the Reynolds number. The aim of this work is to study, using kinetic linear theory, the effect of non-thermal electrons described by a skew Kappa-like function over the excitation of parallel propagating whistler modes associated with the WHFI in a non-collisional, magnetized, solar wind like plasma. We will show that the proposed eVDF reproduces the main field-aligned features of the eVDF as observed in the solar wind, potentially allowing simpler models of solar wind electrons modeled as a superposition of two sub-populations. In addition, considering a unified description of halo and strahl electrons may also be relevant for the understanding of the relevance of the electron non-thermal features for the dynamics of the heat transport by the solar wind~\citep{bale2013electron}, and also the kinetic physics governing the halo formation and its relation with the strahl~\citep{vocks2005electron,Horaites2017,horaites2018}.

The paper is organized as follows: in Section~\ref{sec:model} we present our model, analyze the skew Kappa-like function and introduce it as a new distribution function for describing the halo and strahl electron populations. In section \ref{sec:linear} we show the theoretical results of linear kinetic theory for the dispersion tensor of parallel propagating waves, in a plasma where the electron population is modeled by a core and skew kappa-like function. Then, in section~\ref{sec:WHFI_Thresholds} we systematize this analysis in order to obtain the the marginal stability thresholds for this distribution as a function of plasma beta and heat-flux, and present the best fit parameters for these contours. Finally, in Section~\ref{sec:conclusion} we summarize and discuss our results. 
	
%%%%%%%%%%%%%%%%%%%%%%%%% MODEL %%%%%%%%%%%%%%%%%%%%%%%%%%%%%%

\section{ELECTRON DISTRIBUTION: THE CORE-STRAHLO MODEL}
	   \label{sec:model} 
We model the solar wind electrons distribution function $f_e$ as a superposition of two sub populations. 
\begin{equation}
\label{eq_total_dist}
f_e(v_{\perp}, v_{\parallel}) = f_c(v_{\perp}, v_{\parallel})+ f_{s}(v_{\perp}, v_{\parallel}).    
\end{equation}
A bi-Maxwellian distribution ($f_c$) representing the core
\begin{equation}
\label{eq_core_dist}
    f_c(v_{\perp}, v_{\parallel}) = \frac{n_c}{\pi^{3/2}\alpha_{\perp}^2\alpha_{\parallel}} \ \exp\left(-\frac{v_{\perp}^2}{\alpha_{\perp}^2} - \frac{(v_{\parallel}-U_{c})^2}{\alpha_{\parallel}^2}\right),
\end{equation}
and a skew-Kappa function ($f_s$) to describe both, the halo and strahl electrons, that from now on we will name as the \textit{strahlo}, and this representation of solar wind electrons as \textit{the core-strahlo model}. Under this model $f_s$ consists of a Kappa function to which an asymmetry term has been added. Namely,
\begin{equation}
		\label{eq_skd}
f_{s}(v_{\perp}, v_{\parallel}) = n_{s}A_{s}\left[1+ \frac{1}{\kappa_{s} - \frac{3}{2}}\left(\frac{v_{\bot}^2}{\theta_{\bot}^2} + \frac{v_{\parallel}^2}{\theta_{\parallel}^2} + \delta_{s}\left( \frac{v_{\parallel}}{\theta_{\parallel}} - \frac{v_{\parallel}^3}{3\theta_{\parallel}^3}\right)\right)\right]^{-(\kappa_{s}+1)}.
\end{equation}
In both distribution functions, the subscripts $\parallel $ and $\bot$ are with respect to the background magnetic field, and $n_c$, $n_s$, denote the number density of the core and strahlo, respectively. In Eq.~\eqref{eq_core_dist} $\alpha_\bot$, $\alpha_\parallel$ are the thermal speeds, and $U_c$ is the drift of the core. Also, in Eq.~\eqref{eq_skd} $A_{s}$ is a normalization term such that $\int f_{s}\ d\mathbf{v} = n_{s}$, $\theta_{\parallel}$ and $\theta_{\bot}$ are related to the thermal velocities as defined in Eqs.~\eqref{eq_thetapal} and~\eqref{eq_thetaper}, respectively. Also, $\kappa_s$ is a measurement of the deviation of this function from a Maxwellian distribution, and $\delta_s$ controls the field-aligned skewness. Note that when $\delta_s=0$ we recover the well known Kappa distribution \citep{Olbert1968,vasyliunas1968,Scudder1996,maksimovic2005radial,xiao2006whistler,lazar2016interpretation, lazar2017firehose, vinas2017linear}. 

\subsection{Validity of the model}
\label{sub:validity}

Depending on the value of the $\kappa_s$ and $\delta_s$ parameters, Eq.~\eqref{eq_skd} may become negative, complex or multi-valued, which imposes some caveats and limitations to the use of the skew-Kappa for the electron VDF. In particular, for an arbitrary value of $\delta_s$ and $\kappa_s$ there is a particular value $u = v_\parallel/\theta_\parallel$ in which the skew-Kappa distribution diverges following a vertical asymptote. This value corresponds to the real solution of the following equation
\begin{equation}
    \kappa_s-\frac{3}{2} + \frac{v_{\parallel}^2}{\theta_{\parallel}^2} + \delta_{s}\left( \frac{v_{\parallel}}{\theta_{\parallel}} - \frac{v_{\parallel}^3}{3\theta_{\parallel}^3}\right) = 0,
    \label{eq_asymp}
\end{equation}
that always exists for real values of $\kappa_s$ and $\delta_s$. The dependency of $u$ is strong and weak with respect to $\delta_s$ and $\kappa_s$, respectively. For example, for $\delta_s=0.1$, the real solutions of Eq.~\eqref{eq_asymp} are $u \simeq 30.1$ for $\kappa_s=3$, and $u \simeq 30.4$ for $\kappa_s=10$, respectively. Furthermore, in the case of $\delta_s=0.2$, the values are $u \simeq 15.3$ and $u \simeq 15.7$ (in units of the thermal speed of the strahlo) for $\kappa_s=3$ and $\kappa_s=10$, respectively. Moreover, due to the peak of the VDF at $\mathbf{v} \simeq 0$ and the presence of the mentioned asymptote, the distribution always has a local minima at $u_{min}$, with $0<u_{min}<u$, given by the solution of the derivative of Eq.~\eqref{eq_asymp}. Namely,
\begin{equation}
    u_{min} =\frac{v_{\parallel min}}{\theta_\parallel} = \frac{1+\sqrt{1+\delta_s^2}}{\delta_s}\,,
    \label{eq_min}
\end{equation}
a monotonically decreasing function of $\delta_s$, with $u_{min} \simeq 20.0$ for $\delta_s=0.1$, and $u_{min} \simeq 10.1$ for $\delta_s=0.2$. Therefore, for an arbitrary value of $\delta_s$ there is a speed regime in which the integrals necessary to build the moments of the VDF or the dispersion relation will present vertical asymptotes, branch cuts and poles so that the analytical continuation of the functions in the complex plane may become a quite complicated task. Even though we believe it may be possible to obtain a bounded reasonable solution, such calculation for any arbitrary parameters is beyond the scope of this article. Therefore, our skew-kappa model requires careful treatment when selecting the $\delta_s$ values.
\begin{figure*}[ht]
			\centering
			\includegraphics[width=0.9\textwidth]{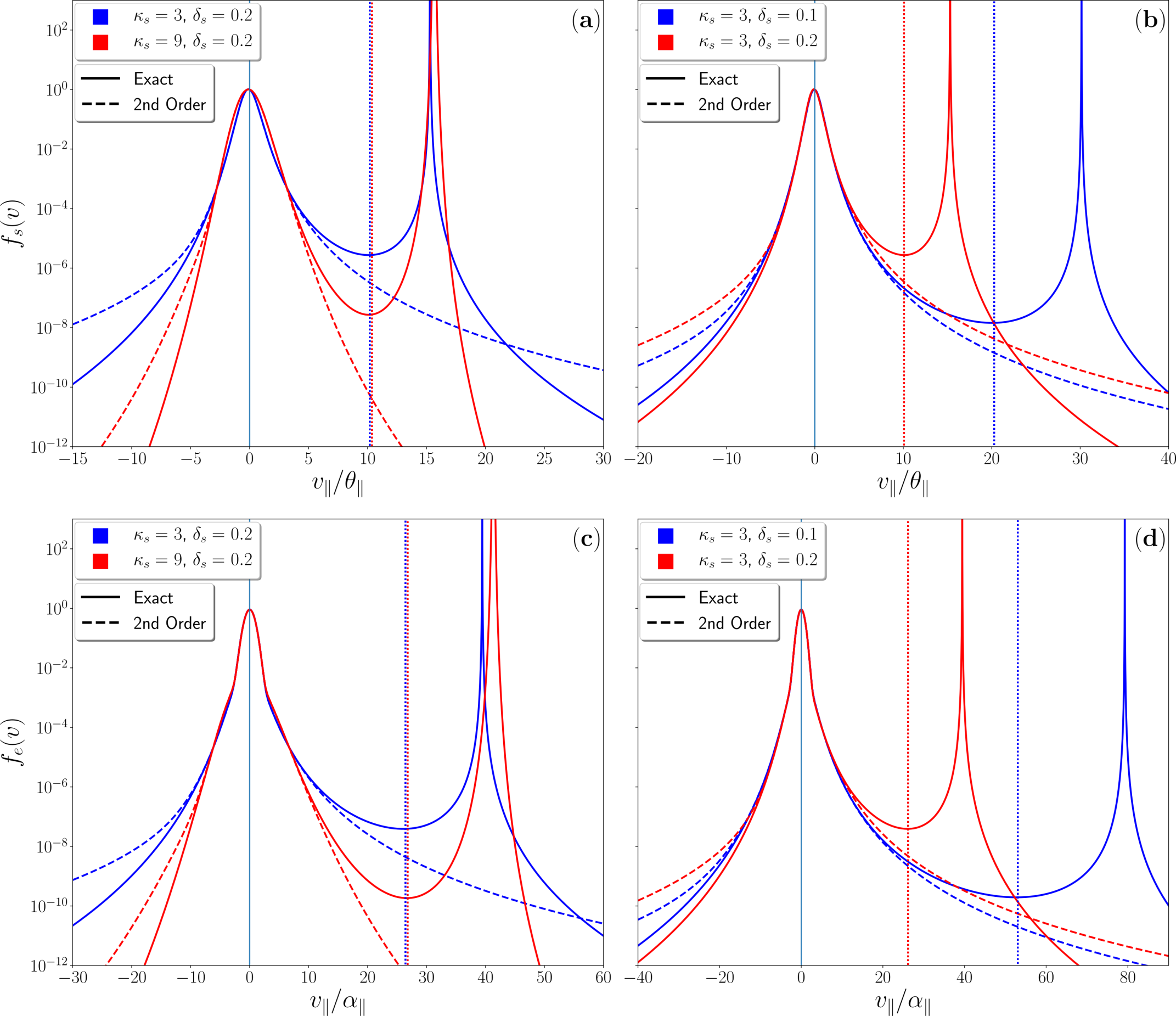}
			\caption{Parallel cuts at $v_{\perp}=0$ of the electron eVDF considering isotropic sub-populations with $n_s/n_e = 0.1$, $T_{\parallel s}/T_{\parallel c} = 7.0$, and different choices of $\kappa_s$ and $\delta_s$. Top and bottom panels
			show the skew-Kappa strahlo and the total electron VDF, respectively.  
			In each panel solid and dashed lines correspond to the exact VDF or a Taylor expansion up to second order in $\delta_s$ respectively. In addition, vertical dotted lines indicate the local minima of the VDF ($v_{min}$) given by Eq.~\ref{eq_min}, and velocities are expressed in unit of the thermal speed of the strahlo (top) or core (bottom).}
			\label{fig_approx}
		\end{figure*}

To avoid these issues, here we apply the heuristic core-strahlo model to situations in which the VDF has small skewness, and the asympote is far away from the main core in units of the thermal speed, so that we can expand all relevant integrals in a finite Taylor series around $\delta_s = 0$. It is important to mention that, since $f_s$ has a vertical asymptote at $u$, such Taylor series is not mathematically possible near $v_\parallel = u \theta_\parallel$. For it to be allowable requires the first derivatives of $f_s$ with respect to $\mathbf v$ must exist, which it does not for those velocity values that are solutions of Eq.~\eqref{eq_asymp}. However, this mathematical problem can be evaded if all relevant features of the distribution are contained at velocities within the $|v_\parallel|/\theta_\parallel < |u_{min}|$ range, i.e., the asymptote of the VDF is far away from the main core. In this case, even though the Taylor series approximation will not be able to mathematically reproduce the exact VDF for all velocity values, calculations based on the approximated version of the VDF in the whole velocity domain will allow analytic calculations, keeping all of relevant physical properties of the skew-Kappa distribution, which subsequently will lead to a direct interpretation of the results and the relevance of each parameter. On the other hand, the general case with arbitrary skewness, when the asymptote may be closer to the main core of the VDF, remains to be solved. In such case the Taylor expansion approach may not be an adequate representation of the VDF near the singularity, and other functional expressions with more attractive properties in the complex plane could be a better option. Under this context, another way to approximate the initial distribution for arbitrary skewness may be the expansion of Eq.~\eqref{eq_skd} on a different base. After a preliminary analysis it seems that the Padé approximant \citep{benderorzag} may be a reasonable procedure for such endeavour, as this approximation does not present new singularities. We will leave this analysis to a future study. From now on we will consider small values of $\delta_s$, such that $\delta^3_s \ll 1$, and we will make use of a Taylor expansion of Eq.~\eqref{eq_skd} up to order $\delta^2_s$ (see Eq.~\eqref{eq_taylor} in Appendix~\ref{Ap_Parm_Mac}).

Figure \ref{fig_approx} shows parallel cuts at $v_{\perp}=0$ of the electron eVDF considering isotropic sub-populations with $n_s/n_e = 0.1$, $T_{\parallel s}/T_{\parallel c} = 7.0$, and different choices of $\kappa_s$ and $\delta_s$. Top and bottom panels show the skew-Kappa strahlo (given by Eq.~\eqref{eq_skd} and the total electron VDF (Eq.~\eqref{eq_total_dist}), respectively, comparing the exact distribution with a Taylor expansion of up to order $\delta^2_s$ as shown in Eq.~\eqref{eq_taylor} in Appendix~\ref{Ap_Parm_Mac}. In addition, vertical dotted lines indicate the value of $v_{min}$ given by Eq.~\ref{eq_min}. From the figure we can see that within the $|v_\parallel| < |v_{min}|$ velocity range the exact and approximated versions of the VDF are mostly the same. In particular, all relevant features of the VDF, such as the skewness and supra-thermal tails, can be clearly observed in both representations inside the $|v_\parallel| < |v_{min}|$ velocity range (as we will see in Section~\ref{sub:core-strahlo-prop}). Thus, as all physical properties of the VDF are covered (shape, moments and dispersion properties), in the small skewness regime ($\delta^3_s \ll 1)$, a second order approximation of electrons following a skew-Kappa distribution given by Eq.~\eqref{eq_skd} can be reasonably represented by the Taylor expansion as shown in Eq.~\eqref{eq_taylor}. In this case all the dispersion functions are reduced to a superposition of standard integrals of the Kappa distribution in $v_\parallel$ similar to the $Q$ integral given by Eq.(5) in~\citet{MaceHellberg1995} or Eq.(12) in~\citet{HellbergMace2002}, a regime already well investigated for integer~\citep{summers1991} or arbitrary~\citep{MaceHellberg1995,HellbergMace2002} values of the $\kappa_s$ parameter. 

To further ascertain the validity of the expansion and the dispersion relation analysis results for the heat-flux instability, we have carried out a comparison of the dispersion properties between a core-halo model based upon drifting Maxwellian distributions and those of the skew-Kappa core-strahlo model, using the same parameters (see Figures~\ref{fig_disp_rel1}b and \ref{fig_disp_rel1}c and their discussion in the next Section). The dispersion results (shown on Figures~\ref{fig_disp_rel1}b and \ref{fig_disp_rel1}c) demonstrate that both model essentially reproduce each other quite well. The real frequencies and growths profiles generated by both model are essentially the same.

Finally, it is worth mentioning that, since the Kappa functions behave as a power-law for large values of the velocity, depending on the value of the $\kappa_s$ parameter, the moments of the distribution may be divergent, which imposes restrictions for $\kappa_s$. In the case of a standard Kappa VDF the pressure is well defined only for $\kappa_s>3/2$. In our case, to have real and finite values of the temperature and heat-flux moments, the values of kappa are restricted to $\kappa_s>5/2$ (see Appendix~\ref{Ap_Parm_Mac} for details). In summary, considering $\delta^3_s \ll 1$ as in the case of this study, up to second order in $\delta_s$ the eVDF is real and positive for all real values of $v_\parallel$, and the integrals in the velocity space share the same poles and branch cuts of Kappa distributions~\citep{MaceHellberg1995,HellbergMace2002}. Consequently, all the moments of the eVDF and also the dielectric tensor of the plasma are well defined within such caveats. 

\subsection{Properties of the core-strahlo model in the small skewness approximation}
\label{sub:core-strahlo-prop}

\begin{figure*}[ht]
			\centering
			\includegraphics[width=0.9\textwidth]{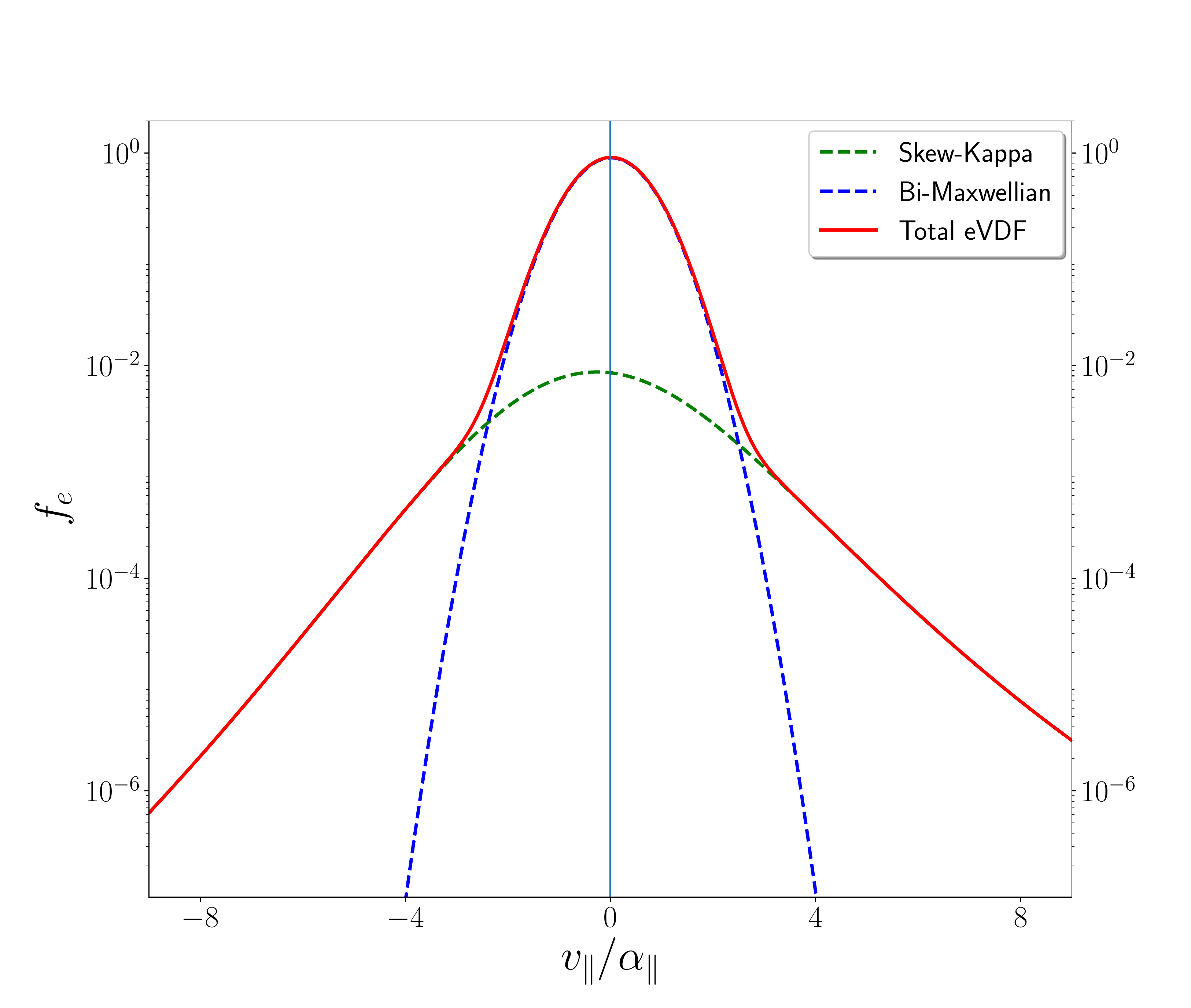}
			\caption{Parallel cuts at $v_{\perp}=0$ of the electron eVDF considering isotropic sub-populations with $n_s/n_e = 0.1$, $T_{\parallel s}/T_{\parallel c} = 7.0$, $\kappa_s=5$, and $\delta_s=0.2$. Blue, green and red curves correspond to core, strahlo, and total eVDF respectively, and velocity is expressed in unit of the thermal speed of the core.}
			\label{fig_components}
		\end{figure*}

\begin{figure*}[ht]
			\centering
			\includegraphics[width=0.9\textwidth]{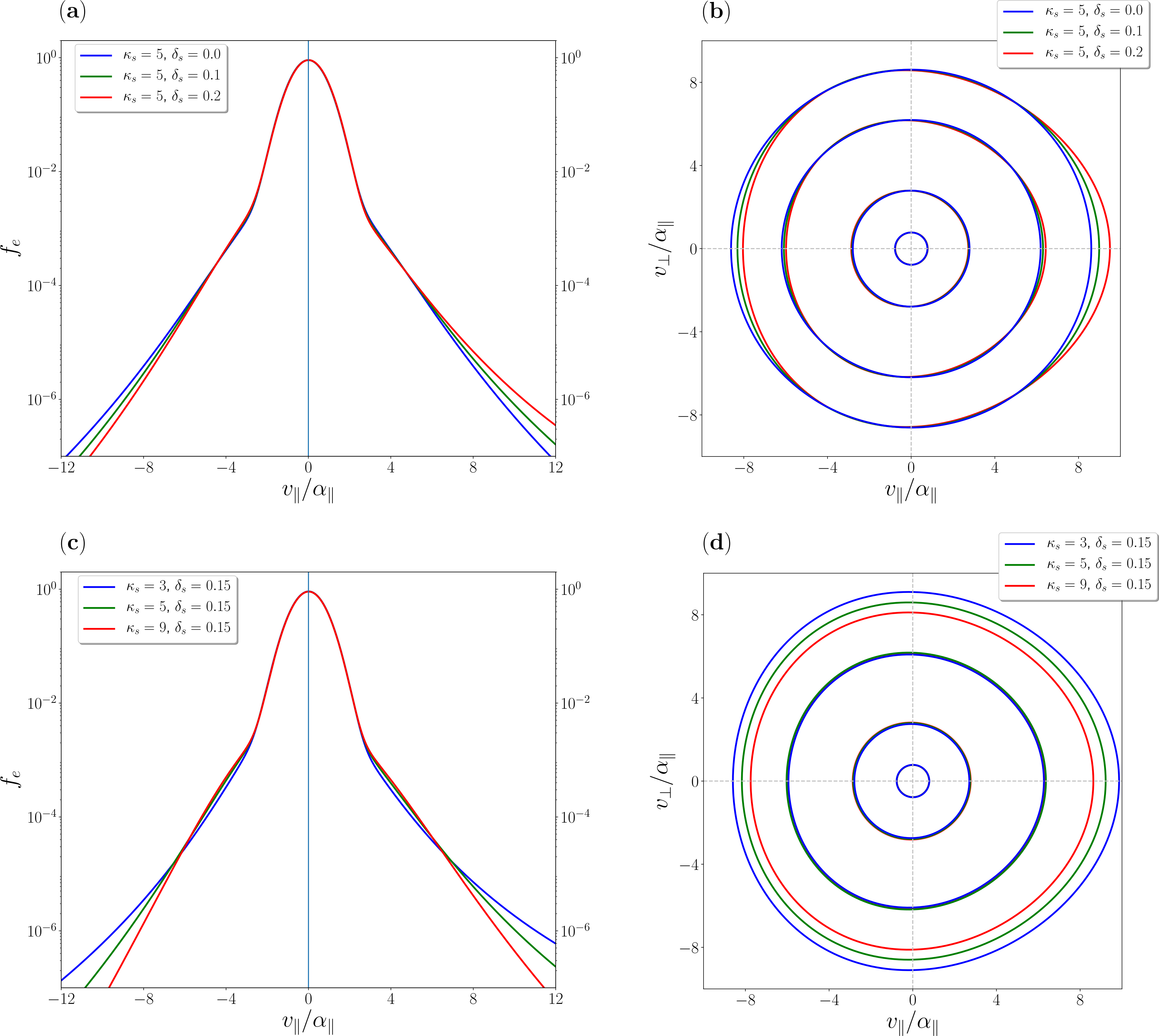}
			\caption{Parallel cuts (left) and contour plots (right) of the eVDF from Eq.~\ref{eq_total_dist}. Top panels consider fixed $\kappa=5$, and different values of the skewness parameter $\delta_s=0$ (blue), $\delta_s=0.1$ (green), and $\delta_s=0.2$ (red); Bottom panels consider fixed skewness ($\delta_s=0.15$), and different kappa values: $\kappa=3$ (blue), $\kappa=5$ (green), and $\kappa=9$ (red). All other parameters are the same as in Figure~\ref{fig_components}. In panels (b) and (d), from innermost to outermost, the levels plotted corresponds to $f(v) = 5\times10^{-1},\ 2\times10^{-3},\ 3\times10^{-5}$ and $2\times10^{-6}$, for all parameter combinations. In all panels parallel and perpendicular velocity components are expressed in unit of the thermal speed of the core ($\alpha_\bot=\alpha_\parallel$).}
			\label{fig_dist}
		\end{figure*}

Even though the core-strahlo model has several free parameters, quasineutrality and zero-current conditions in the ions frame set some relationships between them. In particular, if the ions density is given by $n_p$, to ensure quasineutrality we have $n_e = n_c + n_{s} = n_p$. In other words: 
\begin{equation}
    \frac{n_c}{n_e} + \frac{n_{s}}{n_e} = 1.
\end{equation}
Also, due to the particular shape of the skew-Kappa distribution, for $\delta_s\neq 0$ $f_s$ always has a field-aligned drift $U_s = -\delta_s\theta_\parallel/4$. Thus, the zero-current condition imposes the value of $U_c$ to satisfy
\begin{equation}
    U_{c}\ =\  \frac{n_{s}}{n_c}\ \frac{\delta_s}{4}\theta_{\parallel}.
    \label{eq_Uc}
\end{equation}
Therefore, under this description, in the ions frame there will be a relative drift $\Delta U_\parallel$ between core and strahlo populations given by
$\Delta U_\parallel = \delta_s\,\theta_{\parallel}\,(n_e/n_c)/4$. Note that this relative drift is purely due to the skewness of the strahlo. When the electron distribution has no skewness ($\delta_s=0$), then $f_e$ reduces to a symmetrical distribution with a quasi-thermal core and a non-thermal halo represented by a Kappa distribution~\cite[see e.g][]{pierrard2001core,nieves2008solar}. For more details, full expressions for the macroscopic parameters and the normalization constant of the strahlo distribution function can be found in Appendix \ref{Ap_Parm_Mac}. It is important to recall that these neutrality and quasi-neutrality conditions are restricted to small skewness values. However, as we will show in Figs.~\ref{fig_components}  and~\ref{fig_dist}, our approximation is able to describe thermal and non-thermal electrons in the solar wind, and the parallel cuts of the eVDF have remarkably similar shapes as previously reported using ISEE-1~\cite[see e.g Fig. 1b in][]{ScudderOlbert79} or Wind~\cite[see e.g Fig. 6 in][]{nieves2008solar} data.

To our knowledge, \citet{beck2000application} was the first to propose this type of skew distributions in a study of fully developed hydrodynamic turbulent flows of skew flow velocity distributions via non-extensive statistical mechanics. Under this context the author showed that the asymmetry term, which we have denoted as $\delta_s$, is proportional to Re$^{-1/2}$, where Re is the Reynolds number of the media. This model has been successfully applied to fit data from a turbulent jet experiment~\citep{beck2000application} and environmental atmospheric turbulence~\citep{Rizzo2004}. Here is noteworthy to mention that in both cases the adjusted velocity data lies between $\pm$ 10 thermal speeds as shown in Figure 2 of both studies, and that the obtained skewness parameters are small. In addition, the intrinsic mathematical issues of the skew-Kappa distribution discussed in Section~\ref{sub:validity} can be neglected. Therefore, for both cases the skew-Kappa model represent a useful tool to study the relevance and nature of skew velocity distributions in turbulent flows.  

As turbulence is also present in plasma systems, this suggests that these distributions can be more than an \textit{ad-hoc} function for the solar wind electrons. The $\delta_s$ parameter can potentially be related to microscopic physical processes that allow the particles to exhibit a skew distributions. We strongly believe that this point of view should be further examined, and more rigorous theoretical works studying the underlying physics that allows particle distributions to present this non-thermal feature in plasma systems should be developed. However, such first principle description is beyond the objective of this paper. Here we focus on accepting and using heuristically this skew-Kappa distribution to describe the skewness and high energy tails of the eVDF within the aforementioned caveats. This choice allows us to model the eVDF with less free parameters, as an alternative to the usual ``core-halo-strahl" models \citep{shaaban2019interplay,stverak2009radial, sarfraz2016macroscopic,lopez2019particle}, but at the same time allows us to mimic and study the effect of asymmetry and non-Maxwellian features of the electron population in a solar wind-like plasma, and study their effects on the whistler-heat-flux instability excitation.
		
Figure \ref{fig_components} presents 1D plots at $v_{\perp} = 0 $ of the distribution given by Eq.~\eqref{eq_total_dist} and its two components $f_c$, $f_s$ as a function of the velocity parallel to the mean magnetic field, in unit of the parallel thermal speed of the core. In the figure, blue and green curves represent core and strahlo populations, respectively, and the red curve is the total eVDF. To obtain all of these curves we fixed the density of the non-thermal population (strahlo) to 10\% ($n_s/n_e = 0.1$), and use $T_{\parallel s}/T_{\parallel c} = 7.0$, both of which are solar wind like values \citep{maksimovic2005radial, pierrard2016electron, lazar2020characteristics}. We also considered $\delta_s=0.2$ and $\kappa_s=5$, and isotropic sub-populations i.e $T_{\perp s}/T_{\parallel s} = T_{\perp c}/T_{\parallel c} = 1.0$. The terms $T_{\bot j}$ and $T_{\parallel j}$ correspond to the perpendicular and parallel temperature of population $j$ with respect to the background magnetic field. For the figure we can clearly see that the electron distribution (\ref{eq_skd}) is asymmetric with respect to the $v_{\parallel} =0$, and that this model maintains typical characteristics of solar wind electrons; a Kappa function with enhanced tails and a narrower Maxwellian core for lower energies. Additionally, we can also see that a positive value of the skewness parameter ($\delta_s>0$) enhances the skew-Kappa and the total eVDF to the right such that the total skewness of the distribution is positive (along the field lines).  

Further, Figure \ref{fig_dist} shows 1D plots at $v_{\perp} = 0 $ (left) and contour plots (right) of total distribution \ref{eq_total_dist} for different values of parameters $\delta_s$ (top) and $\kappa_s$ (bottom). In panels \ref{fig_dist}(a) and \ref{fig_dist}(b) we show how the distribution changes for three different values of $\delta_s$ parameter and a fixed value of $\kappa_s = 5.0$. It is clear in both plots that the strahlo loses its symmetry compared to a Kappa function ($\delta_s=0$ case), a feature that is more evident in the outermost contours. Also, we can see that for higher $\delta_s$ values, the more skew is the distribution. In panels \ref{fig_dist}(c) and \ref{fig_dist}(d) we show how the distribution changes for three different values of $\kappa_s$ and fixed $\delta_s = 0.15$. In panel (c) we see that when we increase $\kappa_s$, the high energy tails diminish. This feature is inherited from Kappa distributions, which are reduced to Maxwellian functions in the limit $\kappa \rightarrow \infty$. However, unlike Kappas, skew-Kappa distributions never reduced to Maxwellian distributions because they maintain the skewness for all kappa values when $\delta_s\neq 0$. In panel \ref{fig_dist}(d) we see that the outer contour seem to shrink proportionally as $\kappa_s$ decreases while the core does not change, so that the overall contours' shape appears to remain the same. Therefore, this feature suggests that $\kappa_s$ does not alter the distribution symmetry. 

In summary, the combination of Maxwellian and skew-Kappa distributions can model three important non-thermal features observed in the eVDF in the solar wind, namely, quasi-thermal core, enhanced tails, and skewness. Therefore, its use may allow simpler solar wind models, where electrons are modeled as the superposition of core and strahlo, where the distribution skewness is controlled by only one parameter. This field-aligned skewness provides the energy for the excitation of the WHFI, on which we focus the analysis in this work.

%%%%%%%%%%%%%%%%%%%%%%%%% LINEAR THEORY %%%%%%%%%%

\section{LINEAR THEORY AND DISPERSION RELATION}
\label{sec:linear}	

We use linear kinetic theory to derive the dispersion relation of wave modes that can propagate in a magnetized, non-collisional, and initially uniform plasma. We perform this calculation in order to analyze the stability of the whistler mode associated with WHFI in a solar wind like plasma, where the core-strahlo model is used to describe the electron population. To obtain the dispersion relation we linearize the Vlasov-Maxwell system of equations. This is a well-known method \citep{stix1962theory,krall1973principles} and assumes that the small amplitude perturbations of the relevant quantities are plane waves, allowing the Vlasov-Maxwell system to be rewritten in the form: 
		\begin{equation}
		\label{eq_disp_rel}
		\mathcal{D}(\omega, \mathbf{k}, f_{j})\ \mathbf{\cdot }\ \mathbf{E_k} = 0
		\end{equation}
where $\mathbf{E_k}$ is the complex amplitude of the electric field perturbation and $\mathcal{D}(\omega, \mathbf{k}, f_{j})$ is the dispersion tensor (which is associated with the dielectric tensor of the plasma). This tensor depends on the wave vector $\mathbf{k}$, the complex wave frequency $\omega = \omega_r + i \gamma$ and the background distribution functions of the $j$ species composing the plasma $f_{j}$. The dispersion relation, $\omega = \omega(\mathbf{k})$ is determined by the condition $|\mathcal{D}(\omega, \mathbf{k},f_{j})|=0$, so that Eq. \eqref{eq_disp_rel} has non-trivial solutions for $\mathbf{E_k}$.
		
As a first approximation to the problem, we focus our attention on wave modes that propagate parallel to the background magnetic field $\boldsymbol{B_0}= B_0\hat{z}$, so that $\mathbf{k} = k\hat{z}$. We make this restriction because the mathematical analysis is greatly simplified compared to the oblique case and because previous works have shown that the field-aligned WHFI has larger growth rates \citep{gary1975heat} than the oblique case. We use the core-strahlo distribution given by Eq.~\eqref{eq_total_dist} as the background distribution $f_{e}$ for the electrons. As already mentioned, to perform the integrals involved, we assume that the electron skewness is small, i.e., $\delta^3_s \ll 1$. This approximation allows us to obtain expression for the dispersion tensor elements $D_i = D_i(\omega, k, pp)$ up to second order in $\delta_s$ for the parallel propagating modes. These elements depend on the wavenumber $k$, the wave frequency $\omega$ and the macroscopic parameters of the initial distribution functions (number density, temperature, etc) here denoted as a whole by \textit{pp}. Furthermore, as the distribution is a superposition of Maxwellian and skew-Kappa, the elements of $D_i$ depend on the Fried and Conte plasma dispersion function $Z(\xi)$~\citep{zeta}, and also on the modified dispersion function $Z_{\kappa}(\xi)$~\citep{HellbergMace2002,vinas2015,moya2020towards}. Full expressions of each element of the dispersion tensor can be found in Appendix \ref{Ap_Disp_Tens}. Here we describe the results obtained in the analysis of the excitation of the parallel propagating whistler mode associated with the heat-flux instability.

To obtain the linear properties of the WHFI, we solve the complex dispersion relation using our own developed dispersion solver. We use the core-strahlo distribution (\ref{eq_total_dist}) to model the eVDF, and the proton population is described by an isotropic Maxwellian such that quasi-neutrality and zero-current conditions are both fulfilled. Throughout this analysis we denote proton and electrons parameters with sub-indexes $p$ and $e$ respectively. We fix the proton distribution so that $\beta_{\parallel p} = 0.1$, and $T_{\bot p}/T_{\parallel p} = 1.0$, where $\beta_{\parallel j}= 8\pi n_j k_B T_{\parallel j}/B^2_0$ corresponds to the parallel plasma beta of the population $j$. Also, $k_B$ is the Boltzmann constant, and $n_j$ is the species number density. For the electrons, we fix the anisotropy of both components (core and strahlo) equal to one, i.e, $T_{\bot s}/T_{\parallel s}$ =  $T_{\bot c}/T_{\parallel c} = 1.0$ so that there is no free energy associated with the anisotropy of the eVDF. We also fix the density of the strahlo population to 10\%. With this selection of parameters for protons and electrons, the only relevant non-thermal features of the electron distribution throughout this work, are enhanced tails represented by the $\kappa_s$ parameter, and the skewness represented by $\delta_s$. We perform the stability analysis of the WHFI for different values of $\delta_s,\ \kappa_s$, $\beta_{\parallel s}$ and $T_{\parallel s}/T_{\parallel c}$, to study how the dispersion relation depends on these parameters. Further, as in the solar wind at 1 AU from the Sun, in all our calculations we have fixed the ratio between the electron plasma frequency ($\omega_{pe}$) and gyrofrequency ($\Omega_e$) to $\omega_{pe}/|\Omega_e| = 200$.
		\begin{figure*}[ht]
			\centering
			\includegraphics[width=0.99\textwidth]{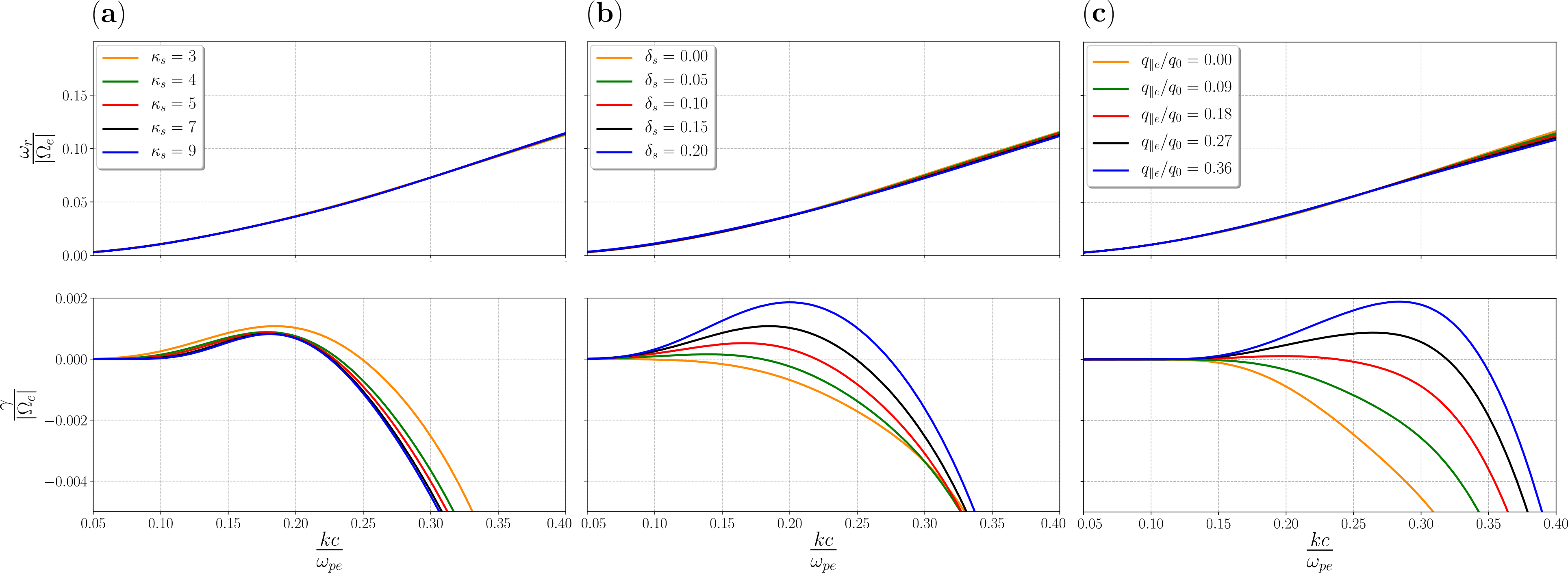}   
    		\caption{Real (top) and and imaginary (bottom) parts of the dispersion relation for the whistler mode for: (a) $\delta_s = 0.15$ and different $\kappa_s$ values; (b) $\kappa_s =3.0$ and different $\delta_s$ values; (c) a core-halo model composed by two drifting Maxwellian with the same heat-flux moment as shown in panel (b) for each value of $\delta_s$. In all cases we set the anisotropy for the electron core and strahlo (or core and halo) equal to one, $n_s/n_e =0.1$, $\beta_{\parallel s} = 1.0$, and $T_{\parallel s}/T_{\parallel c} = T_h/T_c=7.0$.}
			\label{fig_disp_rel1}
		\end{figure*}

Figure \ref{fig_disp_rel1}(a) and \ref{fig_disp_rel1}(b) shows the real (top) and imaginary (bottom) parts of the frequency of the whistler mode for different values of $\kappa_s$ and $\delta_s$, respectively, fixing $\beta_{\parallel s}=1.0$, and strahlo-to-core parallel temperature ratio $T_{\parallel s}/T_{\parallel c} = 7.0$~\citep{maksimovic2005radial, pierrard2016electron, lazar2020characteristics}. Frequency and wavenumber are expressed in units of the electron gyrofrequency, $\Omega_e$, and electron inertial length, $c/|\omega_{pe}|$, where $c$ is the speed of light. Figure \ref{fig_disp_rel1}(a) shows the dispersion relation of the whistler mode considering different values of $\kappa_s$ and fixed skewness parameter ($\delta_s = 0.15$). We can see that the real frequency remains essentially the same when we modify $\kappa_s$ and the growth rates slightly decrease as $\kappa_s$ increases. The case $\kappa_s = 3.0$ corresponds to the most unstable case. As shown in Figure \ref{fig_dist}(c) and \ref{fig_dist}(d), $\kappa_s$ does not control the symmetry of the distribution. Therefore, it is also reasonable the wave stability to slightly depend on $\kappa_s$. Regarding the dependence on the skewness parameter, Figure \ref{fig_disp_rel1}(b) shows the dispersion relation of the whistler mode for different values of $\delta_s$, fixing $\kappa_s = 3.0$. We can see that, in the wavenumber range shown, the real part of the frequency does not change considerably when we increase $\delta_s$. The imaginary part, however, depends more strongly on this parameter, and the wave becomes more unstable as $\delta_s$ increases. Also, the wavenumber range in which the mode is unstable widens, and the $k$ value corresponding to the maximum growth rate also increases with increasing $\delta_s$. Further, note that this relation is non-linear. For example, when $\delta_s=0.1$ the maximum growth rate reaches a value $\gamma_{\rm{max}} \sim 5\times10^{-4}\,|\Omega_e|$, whereas for $\delta_s = 0.2$ the maximum growth rate is $\gamma_{\rm{max}} \sim 2\times 10^{-3}\,|\Omega_e|$. This behavior is expected because $\delta_s$ represents a measurement of the system's free energy associated with the distribution skewness. As $\delta_s$ increases, the more skew the distribution becomes, as we saw in Figure \ref{fig_dist}(a) and \ref{fig_dist}(b). Therefore, it is also expected the relation between the maximum growth rate of the WHFI and the skewness parameter to be non-linear. 
		\begin{figure*}[ht]
			\centering
			\includegraphics[width=0.9\textwidth]{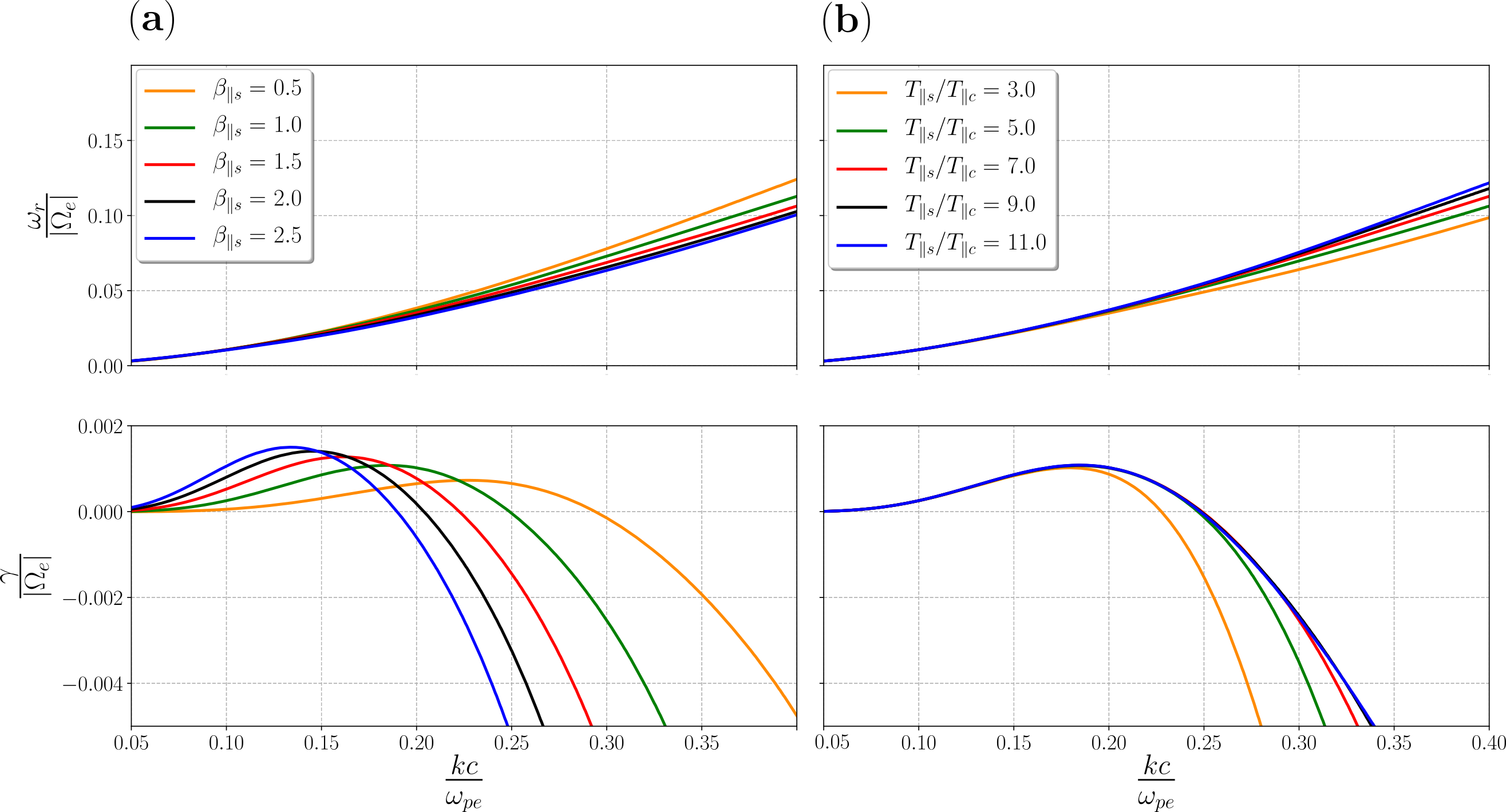}
			\caption{Real (top) and and imaginary (bottom) parts of the dispersion relation for the whistler mode for: (a) $T_{\parallel s}/T_{\parallel c}=7.0$ and different $\beta_{\parallel s}$ values; (b) $\beta_{\parallel s} = 1.0$ and different $T_{\parallel s}/T_{\parallel c}$ values. In all cases we set the anisotropy for the electron core and strahlo equal to one, $n_s/n_e=0.1$, $\kappa_s = 3.0$ and $\delta_s = 0.15$ for the calculations}
			\label{fig_disp_rel2}
		\end{figure*}

Additionally, Figure \ref{fig_disp_rel1}(c) shows similar plots but considering two drifting isotropic Maxwellians given by Eq.~\eqref{eq_core_dist}, so that electrons follow a current-free core-halo model as in~\citet{gary1994whistler}. Under this model, the normalized electron $q_{\parallel e}/q_0$ heat-flux along the magnetic field is giving by $q_{\parallel e}/q_0 \simeq (5/3)(n_c n_h/n^2_e)(\Delta U_{ch}/\alpha_c)(T_h/T_c-1)$, where $T_h$ and $T_c$ are the temperature of the halo and core, respectively, and $\Delta U_{ch}$ is the relative drift between core and halo~\citep{gary1994whistler}. As a comparison with the core-strahlo model, we consider core and halo with the same density, temperatures and plasma beta as core and strahlo. We also select $\Delta U_{ch}$ such that for each $\delta_s$ value shown in Figure \ref{fig_disp_rel1}(b) both models have the same heat-flux moment (we will present more details about the heat-flux of the core-strahlo model in Section~\ref{sec:stability_analysis}). Comparing Figures \ref{fig_disp_rel1}(b) and \ref{fig_disp_rel1}(c), we can see that the dispersion relation obtained by using the skew-Kappa core-strahlo or Maxwellian core-halo models are qualitatively the same. Both models produce the same real part of the dispersion relation, and are similarly unstable to the WHFI. Nevertheless, differences do exist. In particular, for the same level of heat-flux moment the maximum growth of the Maxwellian core-halo model is shifted to larger wavenumber compared to the core-strahlo model. The slight differences between both results are due to the fact that, they are based on different mathematical functions, that have different shapes and velocity gradients (i.e. the dispersion relation depends on these gradients) in the valid domain, and perhaps also, due to the Taylor expansion to second order on the skewness parameter (not present in a Maxwellian description of the plasma). Furthermore, and to the best of our knowledge, these slight differences in the dispersion profiles between the two models have been postulated before e.g. by \citet{abraham1977whistler,Abraham-Shrauner1977b} \citep[and probably extend back to the early work of Bernstein modes;][]{Bernstein1958} in application to whistler and electromagnetic ion cyclotron waves in the solar wind, that indicated that wave dispersion characteristics are not only dependent on the physical moment parameters (e.g., density, temperature, drifts, heat-flux, etc.) but that they also depend on the shape of the distribution. In this case the differences lay on the lack of supra-thermal tails in the Maxwellian model or the fact that in the core-strahlo model the source of asymmetry is strongly dominated by the skewness parameter, not present in the Maxwellian core-halo approach. However, as shown by Figures \ref{fig_disp_rel1}(b) and \ref{fig_disp_rel1}(c) these are minor differences. Both models can adequately describe the WHFI in the small skewness regime. We can definitely conclude that the results of the core-strahlo model are a very good approximation to the heat-flux instability problem, since reproduces similar behavior as previously reported, reinforcing the validity of the model (see Section~\ref{sub:validity}).

Finally, Figure \ref{fig_disp_rel2} shows again the normalized real and imaginary frequencies (top and bottom respectively) of the whistler mode for different values of $\beta_{\parallel s}$ and $T_{\parallel s}/T_{\parallel c}$, for fixed $\delta_s = 0.15$ and $\kappa_s = 3.0$. Figure \ref{fig_disp_rel2}(a) shows the dispersion relation of the whistler mode for different values of $\beta_{\parallel s}$, and fixed $T_{\parallel s}/T_{\parallel c} = 7.0$. From the figure we can see that in this case the real part of the frequency slightly decreases as $\beta_{\parallel e}$ increases. On the other hand, as expected the imaginary frequency depends more strongly on the plasma beta. All cases shown in the plot have a range in which the growth rate is positive, so the plasma is unstable to the whistler mode under these conditions, and it is clear that for higher $\beta_{\parallel s}$ values, the maximum growth rate is also higher. Here, however, the wavenumber at which the growth rate crosses the axis from positive to negative values shifts to the left as $\beta_{\parallel s}$ increases; i.e., the range in which the growth rate becomes positive narrows. Therefore, as a general rule, we can say that for low values of $k$, the growth rates increase with $\beta_{\parallel s}$ and the amplitude of the waves grows faster. At the same time, as $\beta_{\parallel s}$ increases, the wave becomes stable for lower values of the wavenumber. This behavior is also expected since for higher $\beta$ values the less magnetized the plasma is, meaning that the plasma is more susceptible to destabilize due to electromagnetic fluctuations~\citep{vinas2015,moya2020towards}. Further, Figure \ref{fig_disp_rel2}(b) shows the solutions of the dispersion relation for $\beta_{\parallel s}= 1.0$ and different values of the strahlo-to-core ratio $T_{\parallel s}/T_{\parallel c}$. From the figure we can see that in this case, the real part of the frequency increases with increasing $T_{\parallel s}/T_{\parallel c}$. The growth rate also increases with the strahlo-to-core temperature ratio, a expected behaviour as a larger temperature of the only electron component providing the free energy for the instability should also result on a larger growth rate of the waves. However, the maximum value of the growth rate seems to saturate to $\gamma_{\rm{max}} \sim 10^{-3}\,|\Omega_e|$ at $kc\sim 0.2\, \omega_{pe}$ for $T_{\parallel s}/T_{\parallel c}\gtrsim 5$. This is an interesting result, however, as the main goal of our study is the analysis of the whistler heat-flux instability in the solar wind, from now on we will fix $T_{\parallel s}/T_{\parallel c} = 7$ and focus the analysis on the effect of the $\kappa_s$ and $\delta_s$ parameters.

%%%%%%%%%%%%%%%%%%%%%%%%% EFFECT OF HEAT-FLUX  %%%%%%%%%%

\subsection{The effect of the heat-flux moment on the instability}
\label{sec:stability_analysis}
		
We are particularly interested in understanding how the WHFI contributes to the electron heat-flux regulation through collisionless wave-particle interactions in the solar wind. Thus, we expand our stability analysis and study how the whistler wave changes as we modify the electron field-aligned heat-flux moment of the eVDF. The connection between the parallel electron heat-flux ($q_{\parallel e}$) and the parameters describing the electron distribution function (\ref{eq_total_dist}) can be seen in Eq.~\eqref{eq_hf}. Namely, up to second order in $\delta_s$, $q_{\parallel e}$ is given by:
		\begin{equation}
		\label{eq_hf}
		 q_{\parallel e} = \frac{m_e\ n_{s}\ \theta^3_{\parallel}}{4}\ \delta_s\left[\mu_{s}\Psi_6(\kappa_s) +\Psi_7(\kappa_s) + \frac{1}{4}\frac{\alpha_{\parallel}^2}{\theta_{\parallel}^2}\left(3 + 2\mu_c\right)\right]
		\end{equation}
where $\Psi_6$ and $\Psi_7$ are functions that depend only on $\kappa_s$ and $\mu_j = T_{\perp j}/T_{\parallel j}$ is the anisotropy of population 
$j$ (see in Appendix~\ref{Ap_Parm_Mac} for details). Further, to express the heat-flux as a dimensionless quantity it is customary to normalize $q_{\parallel e}$ to the free-streaming or saturation heat-flux $q_0 = (3/2)\,n_e k_B T_{\parallel c}\, \alpha_\parallel$~\citep[see e.g][]{gary1994whistler} Taking this into consideration, we can write the normalized heat-flux as follows:
	\begin{equation}
		\label{eq_norm_hf}
		\dfrac{q_{\parallel e}}{q_0} = \frac{\delta_s}{3}\frac{n_s}{n_e}\left(\frac{T_{\parallel s}}{T_{\parallel c}}\right)^{\frac{3}{2}}\left[\mu_{s}\Psi_6(\kappa_s) +\Psi_7(\kappa_s) + \frac{1}{4}\frac{T_{\parallel c}}{T_{\parallel s}}\left(3 + 2\mu_c\right)\right]
		\end{equation}
In this expression it is clear that the normalized electron heat-flux increases linearly with $\delta_s$. In this case, (when all other parameters are fixed) as we increase $\delta_s$, $q_{\parallel e}/q_0$ also increase and the plasma becomes more unstable to the whistler mode, just like we saw in figure \ref{fig_disp_rel1}(a). In other words, if the heat-flux increases linearly with $\delta_s$, a larger value of $q_{\parallel e}/q_0$ corresponds to a more skew distribution, and therefore indicates a larger level of free energy to excite the WHFI. In contrast, functions $\Psi_5(\kappa_s)$ and $\Psi_6(\kappa_s)$ indicate that the heat-flux decreases as $\kappa_s$ increases. In this case, as we increase $q_{\parallel e}/q_0$, the stability of the plasma to the whistler mode remains essentially the same, just like figure \ref{fig_disp_rel1}b suggest, i.e. when the increase in heat-flux values is a consequence of changes in $\kappa_s$, such increment is due to the modification of the high energy tails, which are enhanced when $\kappa_s$ decreases but do not changes the symmetry of the eVDF.
			\begin{figure}[ht]
			\centering
			\includegraphics[width=0.9\textwidth]{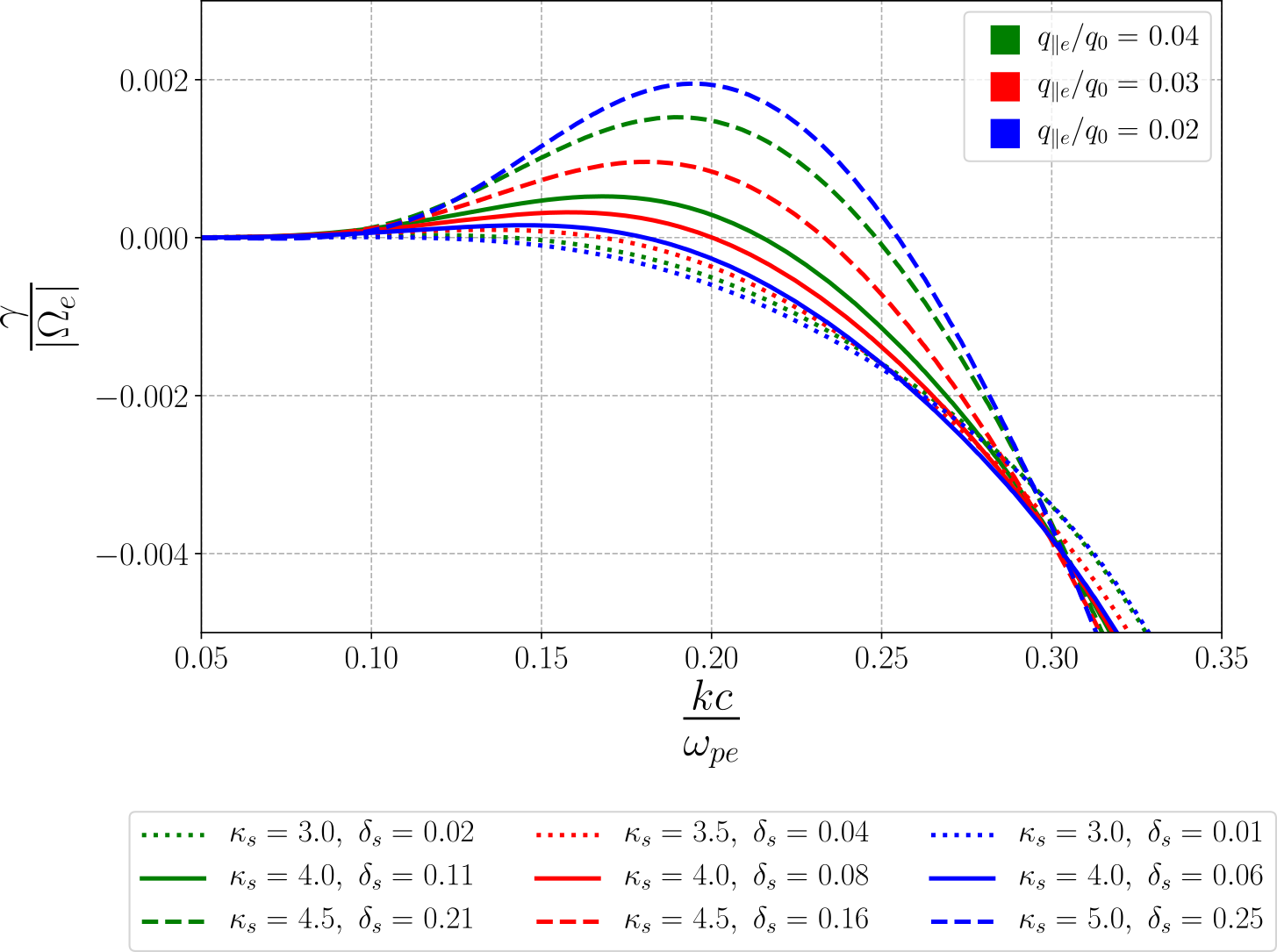}
			\caption{Growth rates of the whistler mode for $\beta_{\parallel s} =1.0$, $\mu_c = \mu_s = 1.0$ and $T_{\parallel s}/T_{\parallel c} = 7$, $n_s/n_e = 0.1$, and different normalized heat-flux values: $q_{\parallel e}/q_0=0.02$ (blue lines), $q_{\parallel e}/q_0=0.03$ (red lines) and $q_{\parallel e}/q_0=0.04$ (green lines). As the same heat-flux can be achieved by multiple combinations of electron parameters, for each value of the heat-flux the growth rates have been derived using 3 distinct ($\kappa_s,\delta_s$) combinations, differentiated by line style (solid, dashed and dotted lines).}
			\label{fig_gr}
		\end{figure} 

To further analyze the behavior of the heat-flux parameter, in Figure \ref{fig_gr} we plot the normalized growth rates $\gamma/|\Omega_e|$ of the whistler mode as a function of the normalized wave number for different values of the initial normalized electron heat-flux $q_{\parallel e}/q_0$. We calculate this parameter using different combinations of $\kappa_s$ and $\delta_s$, and fixing $\beta_{\parallel s} =1.0$ and $T_{\parallel s}/T_{\parallel c} = 7$. Again, we considered isotropic electron populations ($\mu_c =\mu_s=1.0$) and a 10\% density for the strahlo ($n_s/n_e=0.1$). Blue lines correspond to $q_e/q_0 = 0.02$, red lines to $q_e/q_0 = 0.03$ and green lines to $q_e/q_0 = 0.04$. The line styles differentiate the combinations of parameters used. We see that for the same value of $q_e/q_0$, the stability of whistler mode changes depending on the chosen parameters. In this plot, the combinations with higher $\delta_s$ (dashed lines) are always more unstable to this mode: the growth rates are positive in a wider wavenumber range, and the maximum growth rate is higher as well. Another thing that should be noticed is that for higher $\kappa_s$ values higher $\delta_s$ values are needed to achieve the same heat-flux value. 
	
Accordingly, for a fixed value of $\kappa_s$, the heat-flux parameter is a direct measurement of the distribution function's skewness and hence, of the plasma stability (see solid lines in Figure~\ref{fig_gr}). In this case, we can safely say that the higher the initial heat-flux value, the more unstable the whistler mode will be. In contrast, for a fixed $\delta_s$, we cannot make the same straightforward association between the heat-flux and the plasma stability since changing the heat-flux value will not necessarily affect the stability of the whistler mode. Things get more interesting when we allow the variation of both parameters in calculating the initial heat-flux, because the same value of $q_e/q_0$ can be achieved using different combinations of $\kappa_s$ and $\delta_s$. Since only the latter parameter significantly impacts the distribution's skewness, different combinations will have different stability for the whistler mode. In other words, systems whose distributions have different levels of asymmetry and, therefore, different stability to the whistler mode, can have the same heat-flux value. Hence, the heat-flux parameter can no longer be a direct measurement of this non-thermal feature (the asymmetry of the eVDF), which gives the plasma the free energy to radiate electromagnetic waves. In consequence, it is not definitive to assure that higher heat-flux values represent more unstable states.

%%%%%%%%%%%%%%%%%%%%%%%%% INSTABILITY THRESHOLDS  %%%%%%%%%%
	
\section{WHFI INSTABILITY THRESHOLDS}	
\label{sec:WHFI_Thresholds}

		\begin{figure}[ht]
			\centering
			\includegraphics[width=0.9\textwidth]{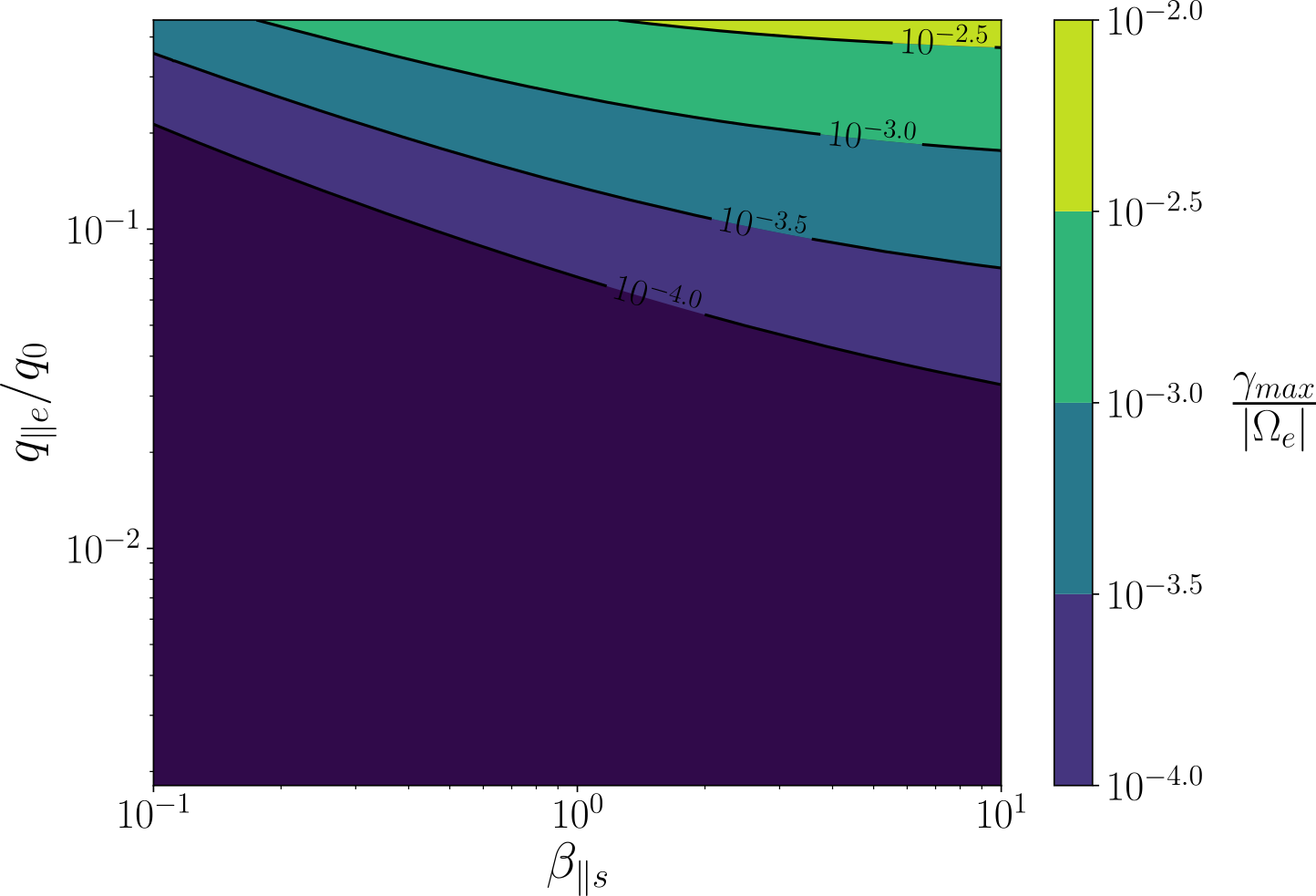}
			\caption{Maximum growth rate of the WHFI, normalized to the electron gyrofrequency $\gamma_{max}/|\Omega_e|$ for $\kappa_s =3.0$, as function of the strahlo parallel beta $\beta_{\parallel s}$ and normalized parallel electron heat-flux $q_{\parallel e}$. The contours shown corresponds to the thresholds $\gamma_{max}/|\Omega_e| = 10^{-2.5},\ 10^{-3.0},\ 10^{-3.5},\ 10^{-4.0}$, and all calculations were performed using $\mu_c = \mu_s = 1.0$, $T_{\parallel s}/T_{\parallel c} = 7$, and $n_s/n_e = 0.1$.}
			\label{fig_contour1k}
		\end{figure}
As mentioned, one of the goals of this research is the understanding about under which plasma conditions the WHFI develops, and also to contrast the theoretical predictions with observational data. To do so, in this section, we systematize our analysis on the excitation of the WHFI, and calculate the instability thresholds for this wave mode in the $q_{\parallel e}/q_0$ vs $ \beta_{\parallel s}$ space. We calculate the normalized maximum growth rate of the WHFI, $\gamma_{max}/|\Omega_e|$, as a function of the normalized electron heat-flux $q_{\parallel e}/q_0$ and the electron beta parameter $\beta_{\parallel s}$. Regarding the macroscopic plasma parameters, for protons we consider the same parameters as mentioned in previous section. For the electrons, as in the previous section we consider isotropic populations ($T_{\perp s}/T_{\parallel s} = T_{\perp c}/T_{\parallel c} = 1.0$) and a density of 10\% for the strahlo population ($n_s/n_e=0.1$). We also use $\kappa_s = 3.0$ and $T_{\parallel s}/T_{\parallel c} = 7.0$.  With this choice of parameters, $q_{\parallel e}/q_0$ is a direct measure of the distribution skewness, as can be seen in Eq. \eqref{eq_norm_hf}. Figure \ref{fig_contour1k} shows a contour plot of the maximum growth rate of the WHFI for $0.1 \le \beta_{\parallel s} \le 10$, and $1.8\times 10^{-3} \le q_{\parallel e}/q_0 \le 0.45$, which following Eq.~\eqref{eq_norm_hf} corresponds roughly to $0.001 \le \delta_e \le 0.25$. From the figure we can see that the general behavior for $\gamma_{max}$ is to increase to the right and upwards in the plot. i.e., as expected, the waves become more unstable as $\beta_{\parallel s}$ and  $q_{\parallel e}/q_0$ increases. Therefore, we recover the behaviors seen in figures \ref{fig_disp_rel1}(a) and \ref{fig_disp_rel2}(a). 
\begin{table}[htbp]
	 \begin{center}
	        \begin{tabular}{|c||c|c|c|c|}
				\hline
				& $A$& $B$ & $\beta_0$&$\alpha$ \\
				\hline \hline
				$\kappa_e = 3.0$ & 0.150 & 0.132 & 0.049 &  0.636\\ \hline

				$\kappa_e = 4.0$ & 0.032 & 0.033 & 0.042 &  0.683\\ \hline
				
				$\kappa_e = 5.0$ & 0.007 & 0.008 & 0.039 &  0.697\\ \hline
			\end{tabular}
		\caption{Best fit parameters for the $\gamma_{max}/|\Omega_e| =  10^{-3}$ threshold of the whistler heat-flux instability. The curve fitting for this thresholds was performed using the function shown in Eq. (\ref{eq_fit}) for different $\kappa_s$, and fixing $\mu_c = \mu_s = 1.0$ and $T_{\parallel s}/T_{\parallel c} = 7$, $n_s/n_e = 0.1$}
		\label{tb_fit}
	\end{center}
\end{table}
		
\begin{figure}[ht]
	\centering
		\includegraphics[width=0.9\textwidth]{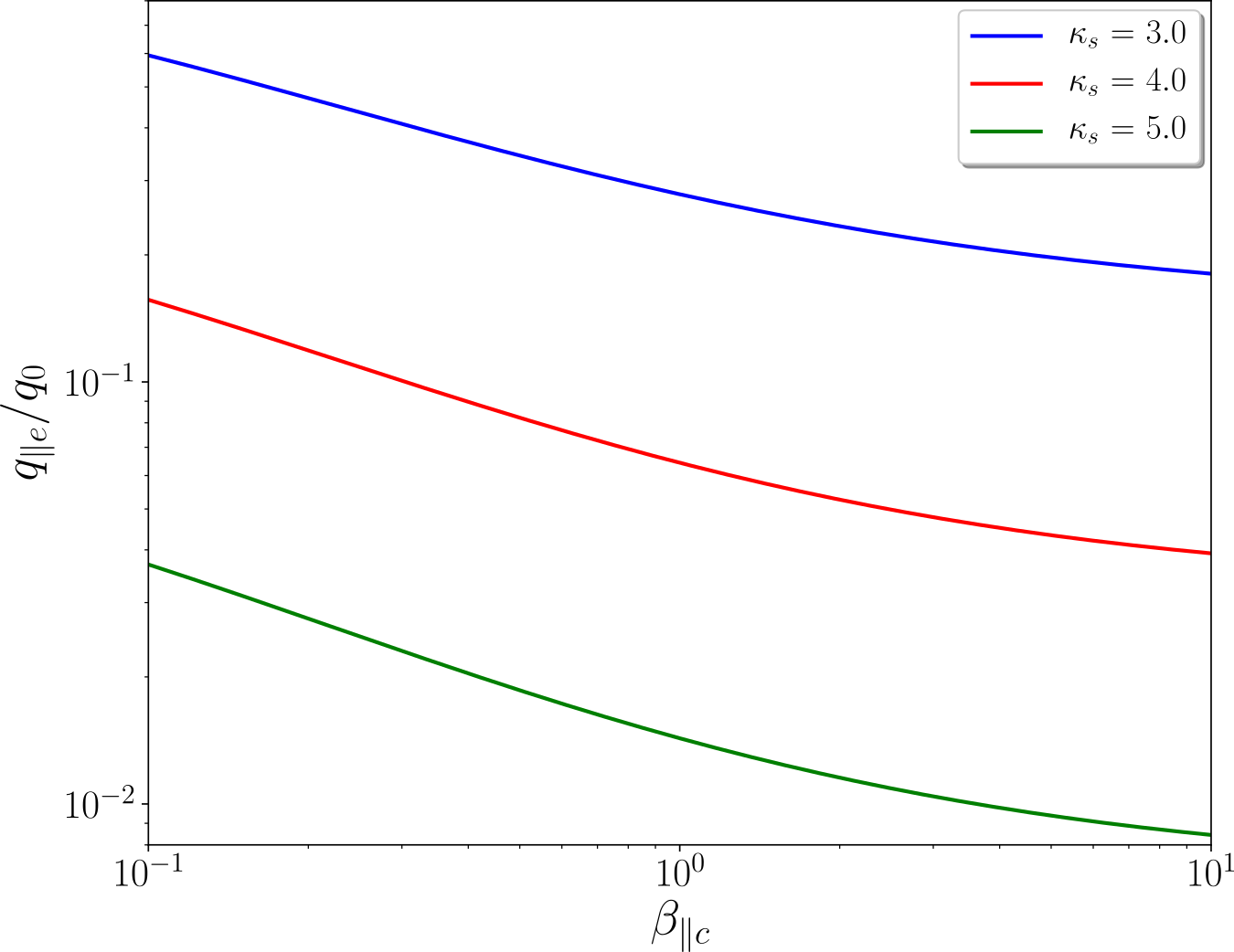}	
		\caption{Fits for the instability thresholds of the whistler heat-flux instability for $\kappa_s = 3$ (blue line), $\kappa_s = 4$ (red line) and $\kappa_s = 5$ (green line). The thresholds shown here correspond to maximum growth rates $\gamma_{max}/|\Omega_e| =  10^{-3}$. All calculations were performed using $\mu_c = \mu_s = 1.0$ and $T_{\parallel s}/T_{\parallel c} = 7$, and $n_s/n_e = 0.1$.}
			\label{fig_contour_all}
\end{figure}

Furthermore, to analyze the effect of $\kappa_s$ on the instability we repeat the calculations for three different $\kappa_s$ values ($3.0$, $4.0$, and $5.0$). In all cases, we calculate the maximum growth rate in the same $\delta_s$ and $\beta_{\parallel s}$ ranges, and use the same proton and electron parameters stated before. However, as the core represents most of solar wind electrons, to facilitate comparisons with observational data we express the growth rates as function of parallel beta of the core. Moreover, for each value of $\kappa_s$ we fit these stability thresholds using a generalized Lorentzian function given by:
\begin{equation}
	\label{eq_fit}
		\frac{q_{\parallel e}}{q_0} = A + \frac{B}{\left(\beta_{\parallel c} + \beta_0\right)^\alpha}\,,
\end{equation}
and we adjust Eq.~\eqref{eq_fit} to the contour $\gamma_{max}/|\Omega_e| = 10^{-3.0}$. In table \ref{tb_fit} we show the parameters $A,\ B,\   \beta_0$ and $\alpha$ of the best fit for every value of $\kappa_s$. We show these parameters to allow easier comparison between these instability thresholds and solar wind observations. Figure \ref{fig_contour_all} shows all of these fits and how the threshold change for different $\kappa_s$ values. Blue, red and green lines correspond to $\kappa_s = 3.0$, $\kappa_s = 4.0$, and $\kappa_s = 5.0$, respectively. From the figure we can see that for a fixed $q_{\parallel e}$ the plasma, predominantly, becomes more unstable as $\kappa_e$ increases. This behavior is consistent with the results shown in Section \ref{sec:stability_analysis}, because as $\kappa_s$ increases, higher $\delta_s$ values are needed in order to achieve the same heat-flux. In other words, as $\kappa_s$ increases, more skew distributions are needed to achieve a given $q_{\parallel e}/q_0$, so more free energy is available in the system to excite waves, which translate into higher growth rates, as we saw in figure \ref{fig_disp_rel1}(a). Therefore, all these results seem to strengthen our previous conclusion that it is not possible to make a direct relation between the heat-flux moment and the stability of the plasma to the WHFI.

%%%%%%%%%%%%%%%%%%%%%%%%% CONCLUSIONS  %%%%%%%%%%	
	
\section{SUMMARY AND CONCLUSIONS}
\label{sec:conclusion}

Using linear kinetic theory, we have performed a stability analysis of the parallel propagating whistler mode associated with the heat-flux instability in a non-collisional, and magnetized plasma, considering parameters typical of the solar wind. Using the Vlasov-Maxwell system, we calculated the dispersion tensor for parallel propagating waves and solved, numerically, the dispersion relation for the whistler mode. To do so, we introduced the \textit{core-strahlo} model, to describe solar wind electrons as a superposition of a quasi-thermal bi-Maxwellian core and the \textit{strahlo}, that represent halo and strahl populations, using a single skew-Kappa distribution function. Observations inspired the usage of this function as it reproduces adequately the high energy tails (measured by the $\kappa_s$ parameter) of the halo and the skewness (measured by the skewness parameter $\delta_s$) introduced by the strahl as seen in the in-situ measurements of the solar wind eVDF. 

Notwithstanding the above, since the skew-Kappa distribution always has a singularity for any finite value of $\delta_s$, the use of the core-strahlo model is restricted to electron distributions exhibiting small skewness (i.e., $\delta_s^3\ll 1$), such that the singularity is far away from the main core of the distribution in units of the thermal speed. As shown in Section~\ref{sub:validity}, in those cases all relevant features of the VDF can be represented through a Taylor series approximation around $\delta_s = 0$, and the singularity can be evaded. Using this model, in the case of small skewness we have studied the effects and sensitivity of different plasma parameters over the excitation of the whistler-heat-flux instability, such as $\kappa_s$, $\beta_{\parallel s}$, $T_{\parallel s}/T_{\parallel c}$ and the skewness parameter $\delta_s$. We focused our attention on the effect that the macroscopic parameter $q_{\parallel e}$ has in the mode's stability since this is the parameter that has been customarily used as a measurement of the eVDF skewness, which is the non-thermal feature in the system that provides the energy for the excitation of the WHFI. We also characterized the $\beta$ dependent marginal stability thresholds as a function of the parallel electron heat-flux parameter, and present threshold conditions for the instability that can be modeled to compare with observational data.

Our results showed that when $\delta_s>0$ the plasma is unstable to the parallel propagating WHFI, and that the growth rates increase with increasing $\delta_s$, which is the parameter that controls the skewness of the eVDF. In addition, we also showed that $\kappa_s$ (the parameter controlling the extent of the high-energy power-law tails of the distribution) has a weak effect on the stability of this mode: as we increase its value, the mode becomes slightly more stable. Furthermore, we presented the analytical expression for the normalized electron heat-flux. For the isotropic case ($\mu_e=1$) we showed that it is not possible to definitely predict how the growth rates will modify as we increase the electron heat-flux. This behavior results from the fact that a given $q_{\parallel e}/q_0$ value can be achieved by multiple $(\kappa_s, \delta_s)$ combinations. Therefore the stability of the whistler mode depends greatly on how $q_{\parallel e}/q_0$ is calculated in terms of $\delta_s$ and $\kappa_s$. Finally, to allow comparison with observations, we presented the best-fit parameters, where a generalized Lorentzian has been used for the curve fitting of these stability thresholds. Considering that in this model only $\delta_s$ controls the distribution skewness, and that high $\delta_s$ values (rather than high $q_e/q_0$ value) are consistent with more unstable states, our results suggest that studies regarding the excitation of the WHFI should be mostly focused on the distribution skewness (a purely kinetic property of the VDF) instead of the heat-flux moment, which is a fluid quantity of the plasma. Under this context, we expect these results to provide a new framework to study the role of the WHFI on the depletion of the field-aligned electron heat-flux below the values predicted by the collisional transport model and clearly observed by \citet{bale2013electron}. Moreover, we also expect our results to motivate the search for new methods to measure or estimate the skewness or asymmetry of the eVDF from observations, rather than the field-aligned heat-flux moment.

That being said, is worth mentioning that a skew distribution can also be unstable to other micro-instabilities. In particular, as recently shown by~\citet{lopez2020alternative,Lopez2020}, electron distribution functions composed by a core and a beam can be unstable to several instabilities. However, the electrostatic instability is the fastest growing mode only when the relative drift between core and beam is larger than the thermal speed. Furthermore, even when the electrostatic mode is faster than the electromagnetic instability, its saturation level is also faster but lower than the electromagnetic. Therefore the electromagnetic mode dominates in the non-linear regime. In the case of our study, by construction the relative drift between core and strahlo is smaller than the thermal speed. Thus, the electrostatic instability may be present, but the instability triggered by the skewness (the heat-flux instability) should dominate. Due to this reason, and also because of the particular interest of the community on the possible role of the WHFI on the regulation of solar wind heat-flux through wave-particle interactions, we have decided to focus only on the whistler heat-flux instability (WHFI). Nevertheless, we acknowledge the possible existence of more instabilities due to the different non-thermal properties of the electron distribution (temperature anisotropy, asymmetry, relative drifts, power-law tails, etc). We believe it is important to address the coexistence and interplay of these instabilities, and we expect to perform such analysis in a subsequent study.

As previously mentioned, to our knowledge the skew-Kappa function has never been used in a space plasma context. In the original derivation of this type of distribution in the context of turbulent flows, \citet{beck2000application} showed that the inverse of the asymmetry term $1/\delta_s$, is proportional to the square root of the Reynolds number (Re$^{1/2})$. Also, a quick calculation shows that Re is inversely proportional to the Knudsen number Kn (Re $\propto$ Kn$^{-1}$). As Kn is directly related with the heat-flux transport in a collisional plasma~\citep{spitzer1953transport,bale2013electron}, this relation between $\delta_s$ and Kn suggests that $\delta_s$ can potentially be related to parameters relevant to the turbulence phenomenon and plasma collisionality, making a possible connection between the kinetic properties of the plasma (the skewness) and a fluid description of the media (the Knudsen number). Therefore, the core-strahlo model based on the skew-Kappa distribution can be more than an ad-hoc representation of solar wind electrons. Even though the detailed explanation of such relation is beyond the scope of these work, we strongly believe that theoretical studies that can link these parameters in plasma systems should be investigated further. For example, as mentioned in~\citet{bale2013electron}, the transition between the collisional Spitzer-H\"arm heat-flux transport and the collisionless regimes occurs at Kn $\sim$ 0.3. Therefore, using Eq.~\ref{eq_norm_hf}, in terms of $\delta_s$ our model predicts that such transition should occur at $\delta_s\sim0.2$ for $\kappa_s=3$, $n_s/n_e =0.1$, $T_{\parallel s}/T_{\parallel c} = 7.0$, and $\mu_c=\mu_s=1$.

In summary, through the introduction of the core-strahlo model we have showed that plasma states with the same initial heat-flux could have different stability to the WHFI. In other words, systems with high $q_{\parallel e}/q_0$ values can be stable enough so that the WHFI cannot modify the electron heat-flux values effectively through wave-particle interactions. Thus, the heat-flux by itself does not seem to be the best indicator of the WHFI. The precise source of the a field-aligned heat-flux instability is the skewness of the distribution, and in the case of the skew-Kappa, such non-thermal feature is clearly represented by the skewness parameter $\delta_s$. The use of the skew-Kappa eVDF allows the distribution skewness to be controlled through just one parameter, and considerably reduces the space of free parameters to analyze. Furthermore, the core-strahlo model represents an alternative approach to the usual functions used to phenomenologically model the eVDF in terms of a superposition of three electron sub populations: core, halo and strahl. This new model to describe the eVDF greatly simplify the study of WHFI, and the role this instability plays in the electron thermal energy transport in the solar wind. Finally, the use of this distribution has a potential theoretical justification that should be explored as $\delta_s$ could be related to the turbulence phenomenon in plasma systems, and also be relevant for the understanding of the relation between the strahl and the formation of the halo during the expansion of the solar wind from the outer solar corona to the heliosphere. Under this context, since the general case with arbitrary skewness remains to be solved, our results suggest that future research about the collisionless heat-flux transport in the solar wind should centered on the distribution skewness (encapsulated here by the $\delta_s$ parameter), rather than just the heat-flux parameter. We expect this theoretical analysis inspired by observations to be relevant and provide valuable insights in the solar wind electron heat-flux regulation debate. Hopefully, the results shown here will be validated with experimental data in lights of the new Parker Solar Probe and Solar Orbiter missions.

\noindent 
\acknowledgements

\noindent We thank the support of ANID, Chile through the  Doctoral National Scholarship N$^{\circ}$21181965 (B.Z.Q.) and FONDECyT grant No. 1191351 (P.S.M.). A. F.- Vinas would like to thank the Catholic University of America/IACS and NASA-GSFC for their support during the development of this work.

%\bibliographystyle{IEEEtran}
%{\small
%\bibliography{poster}}

%%%%%%%%%%%%%%%%%%%%%%%%% APPENDIX  %%%%%%%%%%

\appendix

\section{Macroscopic parameters}
\label{Ap_Parm_Mac}

In this Appendix we show expressions for the relevant macroscopic parameters of the skew-Kappa distribution function $f_s$ shown in Eq. \eqref{eq_skd}. In order to perform the integrals in velocity space involved in these expressions, we assume a small skewness (i.e., $\delta^3_s \ll 1$) and compute these macroscopic parameters using a Taylor expansion of Eq~\eqref{eq_skd} up to second order in $\delta_s$. As usual, in the following equations the subscripts $\parallel$ and $\perp$ are with respect to the background magnetic field.

\begin{enumerate}
    \item \textbf{Taylor expansion of $f_s$}
\begin{equation}
    f_{s}(v_{\perp}, v_{\parallel}) = n_sA_s\left[F_0(v_{\perp}, v_{\parallel}) - \delta_s F_1(v_{\perp}, v_{\parallel}) + \frac{\delta_s^2}{2}F_2(v_{\perp}, v_{\parallel})+\mathcal{O}(\delta_s^3)\right]\,.
    \label{eq_taylor}
\end{equation}
Where: 
\begin{equation*}
    F_0(v_{\perp}, v_{\parallel}) = \left[1+ \frac{1}{\kappa_{s} - \frac{3}{2}}\left(\frac{v_{\bot}^2}{\theta_{\bot}^2} + \frac{v_{\parallel}^2}{\theta_{\parallel}^2}\right)\right]^{-(\kappa_{s}+1)}\,,
\end{equation*} 
\begin{equation*}
    F_1(v_{\perp}, v_{\parallel}) = \left[\frac{\kappa_s+1}{\kappa_s-\frac{3}{2}}\right]\left[ \frac{v_{\parallel}}{\theta_{\parallel}} - \frac{v_{\parallel}^3}{3\theta_{\parallel}^3}\right]\left[1+ \frac{1}{\kappa_{s} - \frac{3}{2}}\left(\frac{v_{\bot}^2}{\theta_{\bot}^2} + \frac{v_{\parallel}^2}{\theta_{\parallel}^2}\right)\right]^{-(\kappa_{s}+2)}
\end{equation*}
and
\begin{equation*}
    F_2(v_{\perp}, v_{\parallel}) = \frac{\delta_s^2}{2}\left[\frac{\kappa_s+1}{\kappa_s-\frac{3}{2}}\right]\left[\frac{\kappa_s+2}{\kappa_s-\frac{3}{2}}\right]\left[ \frac{v_{\parallel}}{\theta_{\parallel}} - \frac{v_{\parallel}^3}{3\theta_{\parallel}^3}\right]^2\left[1+ \frac{1}{\kappa_{s} - \frac{3}{2}}\left(\frac{v_{\bot}^2}{\theta_{\bot}^2} + \frac{v_{\parallel}^2}{\theta_{\parallel}^2}\right)\right]^{-(\kappa_{s}+3)}\,.
\end{equation*}

	\item \textbf{Number density}
	\begin{equation}
	n_s = \int f_s\ d^3v =  n_sA_s\frac{\left[(\kappa_s - 3/2)\pi\right]^{\frac{3}{2}}\theta_{\perp}^2\theta_{\parallel}\Gamma(\kappa_s - 1/2)}{\Gamma(\kappa_s+1)} \left[1+\frac{\delta_s^2}{4}\Psi_1(\kappa_s)\right]\,,
	\end{equation}
where we have defined 
\begin{equation}
\Psi_1(\kappa)=\left(\dfrac{2\kappa - 1}{2\kappa - 3}\right)- \dfrac{7}{12}\,.
\end{equation}	
Therefore, under this regime, the normalization constant $A_s$ in terms of the number density is given by:

	\begin{equation}
	A_s = \dfrac{\Gamma(\kappa_s+1)}{\left[(\kappa_s - 3/2)\pi\right]^{3/2}\theta_{\perp}^2\theta_{\parallel}\Gamma(\kappa_s - 1/2)} \left[1-\frac{\delta_s^2}{4}\Psi_1(\kappa_s)\right]\,.
	\end{equation}
	
	\item \textbf{Parallel bulk velocity}
	\begin{equation}
	U_s = \frac{1}{n_s}\int v_{\parallel}f_s\ d^3v = -\frac{\delta_s}{4}\theta_{\parallel}\,.
	\end{equation}
	
	\item \textbf{Perpendicular temperature}
	\begin{equation}
	T_{\perp s} = \dfrac{m_s}{2n_sk_B}\int v_{\perp}^2f_s\ d^3v = \dfrac{m_s\ \theta_{\perp}^2}{2k_B}\left[1+\dfrac{\delta_s^2}{4}\Psi_2(\kappa_s)\right]\,,
	\label{eq_thetaper}
	\end{equation}
	where 
	\begin{equation}
	\Psi_2(\kappa) = \dfrac{5}{12}\left(\dfrac{2\kappa-3}{2\kappa-5}\right) - \left(\dfrac{2\kappa-1}{2\kappa-3}\right) + \dfrac{7}{12}\,.    
	\end{equation}

	\item \textbf{Parallel temperature}
	
	\begin{equation}
	T_{\parallel s} = \dfrac{m_s}{n_sk_B}\int (v_{\parallel} - U_s)^2f_s\ d^3v = \dfrac{m_s\ \theta_{\parallel}^2}{2k_B}\left[1+\dfrac{\delta_s^2}{4}\Psi_3(\kappa_s)\right]\,,
	\label{eq_thetapal}
	\end{equation}
	where we have defined 
	\begin{equation}
	\Psi_3(\kappa) = \dfrac{35}{12}\left(\dfrac{2\kappa-3}{2\kappa-5}\right) - \left(\dfrac{2\kappa-1}{2\kappa-3}\right) - \dfrac{23}{12}\,.    
	\end{equation}
	\item \textbf{Parallel heat-flux} 
	
	\begin{equation*}
	q_{\parallel s} = \dfrac{1}{2}m_s\int (\vec{v}-\vec{U}_s)^2 (v_{\parallel}-U_s)f_s\ d^3v = \frac{m_s\ n_s\ \theta_{\parallel}^3}{8}\ \delta_s\left[\mu_s\Psi_4(\kappa_s) + \Psi_5(\kappa_s)\right]\,,
	\end{equation*}
	where 
\begin{equation}
    \Psi_4(\kappa) = \left(\dfrac{2\kappa - 3}{2\kappa - 5} \right) -1
\end{equation}
	and
	\begin{equation}
	    \Psi_5(\kappa) = \frac{5}{2}\left(\dfrac{2\kappa - 3}{2\kappa - 5} \right) - \frac{3}{2}\,.
	\end{equation}
    
Thus, the parallel heat-flux for the total electron distribution (\ref{eq_total_dist}) is given by:

		\begin{equation}
		\label{eq_hfA}
		 q_{\parallel e} = \frac{m_e\ n_{s}\ \theta^3_{\parallel}}{4}\ \delta_s\left[\mu_{s}\Psi_6(\kappa_s) +\Psi_7(\kappa_s) + \frac{1}{4}\frac{\alpha_{\parallel}^2}{\theta_{\parallel}^2}\left(3 + 2\mu_c\right)\right],
		\end{equation}
		where 
\begin{equation}
    \Psi_6(\kappa) = \frac{1}{2}\left(\dfrac{2\kappa - 3}{2\kappa - 5} \right) -1
\end{equation}
	and
	\begin{equation}
	    \Psi_7(\kappa) = \frac{5}{4}\left(\dfrac{2\kappa - 3}{2\kappa - 5} \right) - \frac{3}{2}\,.
	\end{equation}

\end{enumerate}

\section{Dispersion tensor}
\label{Ap_Disp_Tens}
%In a non-colissional and initially uniform plasma, immersed in a background magnetic field $\mathbf{B_0} = B_0\hat{z}$, the dispersion tensor $\mathcal{D}$ for parallel propagating modes $\mathbf{k} = k\hat{z}$ takes the form.
% donde las especies que lo componen tienen función de distribución $f_0^s$ in the special case of parallel propagating waves, takes the form

%The dispersion tensor $\mathcal{D}$ depends on the initial distribution functions $f_{0s}$ of all species composing the plasma. The restriction $|\mathcal{D}(\omega, k,f_{0s})|$ determines the relation between the wave frequency $\omega$ and the wave number $\mathbf{k} = k\hat{z}$ for the parallel propagating modes. 

%\begin{eqnarray*}
%	D_1 &=& 1- \frac{kc^2}{\omega^2} + 4\pi\sum_s\chi_1^s\\
%	D_2 &=& 4\pi\sum_s\chi_2^s\\
%	D_3 &=& 1 + 4\pi\sum_s\chi_3^s
%\end{eqnarray*}
In a non-colissional and uniform plasma, immersed in a background magnetic field $\mathbf{B_0} = B_0\hat{z}$, for parallel propagating waves $\mathbf{k} = k\hat{z}$, the dispersion tensor $\mathcal{D}$ can be written in the following form:

\begin{equation}
\label{disp_tens} 
\mathcal{D}(\omega, k, f_{j}) = \begin{pmatrix}
1- \frac{kc^2}{\omega^2} + 4\pi\sum_s\chi_1(f_{j})& 4\pi\sum_s\chi_2(f_{j}) & 0 \\ 
-4\pi\sum_s\chi_2(f_{j})& 1- \frac{kc^2}{\omega^2} + 4\pi\sum_s\chi_1(f_{j}) & 0\\ 
0 & 0 &1 + 4\pi\sum_s\chi_3(f_{j})
\end{pmatrix}
\end{equation}
%The restriction $|\mathcal{D}(\omega, k,f_{0s})|$ determines the relation between the wave frequency $\omega$ and the wave number $\mathbf{k} = k\hat{z}$ for the parallel propagating modes. In expression \ref{disp_tens} we see that the dispersion tensor elements are written in terms of the susceptibilities $\chi_i(f_{0s})$ of the species s and the sum is carry on over all plasma populations. The functional form of $\chi_i(f_{0s})$ depends on the initial distribution function $f_{0s}$ of the species $"s"$. In particular, considering the skew-Kappa function shown in eq. (\ref{eq_skd}) and assuming a small skewness (i.e $\delta_s<<1$), the susceptibilities $\chi_i$, up to second order in $\delta$ take the following form:
The restriction $|\mathcal{D}(\omega, k,f_{j})|$ determines the relation between the wave frequency $\omega=\omega_r + i\gamma$ and the wavenumber $k$ for the parallel propagating modes. In Eq. \eqref{disp_tens} the dispersion tensor elements are written in terms of the susceptibilities $\chi_i(f_{j})$, where the sums are carried on over all species $j$ composing the plasma. The functional form of $\chi_i(f_{j})$ depends on the initial distribution function $f_{j}$ describing population $j$. Considering the Taylor approximation of the skew-Kappa function $f_s$ shown in Eq. \eqref{eq_taylor}, all resonant integrals, relevant for the construction of the dispersion tensor, are reduced to the same integrals necessary to compute to obtain the dispersion relation in a Kappa distributed plasma as already studied by~\citet{summers1991}, \citet{MaceHellberg1995} and~\citet{HellbergMace2002}. Therefore, within the small skewness approximation it is safe to assume that all poles and branch-cuts have been included, and the integrals do not present any other contribution. Thus, up to second order in $\delta_s$ the susceptibilities $\chi_i$ of the strahlo take the following form:
%In particular
\begin{equation}
	\chi_1(f_s) = -\frac{\omega_{ps}^2}{4\pi\omega^2} + \frac{1}{8\pi}\frac{\omega_{ps}^2}{\omega^2}\left(\frac{\theta_{\perp}}{\theta_{\parallel}}\right)^2\sum_{n=-1,1}\left(\Lambda_{n}^0+ \delta_s \Lambda_{n}^1 + \frac{\delta_s^2}{2}\Lambda_n^2\right)\\
\end{equation}
\begin{equation}
	\chi_2(f_s) = \frac{i}{8\pi}\frac{\omega_{ps}^2}{\omega^2}\left(\frac{\theta_{\perp}}{\theta_{\parallel}}\right)^2\sum_{n=-1,1}n\left(\Lambda_{n}^0+ \delta_s \Lambda_{n}^1 + \frac{\delta_s^2}{2}\Lambda_n^2\right)\\
\end{equation}
\begin{equation}
\chi_3(f_s) = \frac{1}{2\pi}\frac{\omega_{ps}^2}{k^2\ \theta_{\parallel}^2}\left(\frac{2\kappa_s - 1}{2\kappa_s - 3}\right)\left(\Lambda_0 + \delta_s \Lambda_1 + \frac{\delta_s^2}{2}\Lambda_2\right)
\end{equation}
where the elements of the expansion are given by:
\begin{equation}
	\Lambda_n^0 = 1 + \varphi_{n\kappa}Z_{\kappa}(\xi_{n\kappa})\\
\end{equation}
\begin{equation}	
	\Lambda_n^1 = \frac{1}{2}\sqrt{\frac{\kappa_s}{\kappa_s - \frac{3}{2}}}(1-\xi_n^2)Z_{\kappa}(\xi_{n\kappa}) + \frac{1}{2}\left(\frac{\varphi_n}{3} -\xi_n\right)
	-\left(\frac{\kappa_s- \frac{1}{2}}{\kappa_s - \frac{3}{2}}\right)\left(1-\frac{\xi_n^2}{3}\right)\varphi_n[1+\bar{\xi}_{n\kappa}Z_{\kappa+1}(\bar{\xi}_{n\kappa})]\\
\end{equation}

\begin{equation}
\begin{split}
\Lambda_n^2 = & \left(\frac{\kappa_s - \frac{1}{2}}{\kappa_s-\frac{3}{2}}\right)\left(\frac{\kappa_s +\frac{1}{2}}{\kappa_s-\frac{3}{2}}\right)\varphi_n\xi_n\left(1-\frac{\xi_n^2}{3}\right)^2[1+\tilde{\xi}_{n\kappa}Z_{\kappa+2}(\tilde{\xi}_{n\kappa})] + \frac{1}{2}\left(\frac{\kappa_s - \frac{1}{2}}{\kappa_s-\frac{3}{2}}\right) + 
\frac{1}{3}\left(\frac{\kappa_s - \frac{1}{2}}{\kappa_s-\frac{3}{2}}\right)\varphi_n\xi_n\left(\frac{\xi_n^2}{6}-1\right)\\ \
&-\left(\frac{\kappa_s - \frac{1}{2}}{\kappa_s-\frac{3}{2}}\right)(1-\xi_n^2)\left(1-\frac{\xi_n^2}{3}\right)[1+\bar{\xi}_{n\kappa}Z_{\kappa+1}(\bar{\xi}_{n\kappa})] - \frac{\xi_n}{6}\left(\xi_n -\frac{\varphi_n}{2}\right) +\frac{1}{8} - \frac{\Psi_1(\kappa_s)}{2}[1 + \varphi_{n\kappa}Z_{\kappa}(\xi_{n\kappa})], 
\end{split}
\end{equation}
for $n =1,-1$, and
\begin{equation}
	\Lambda_0 = 1 + \bar{\xi}_{0\kappa}Z_{\kappa+1}(\bar{\xi}_{0\kappa})
\end{equation}
\begin{equation}
	\Lambda_1 = \frac{1}{2}\sqrt{\frac{\kappa_s+1}{\kappa_s - \frac{3}{2}}}(1-\xi_0^2)Z_{\kappa+1}(\bar{\xi}_{0\kappa}) -\frac{\xi_0}{3} -\left(\frac{\kappa_s+ \frac{1}{2}}{\kappa_s - \frac{3}{2}}\right)\left(1-\frac{\xi_0^2}{3}\right)\xi_0[1+\tilde{\xi}_{0\kappa}Z_{\kappa+2}(\tilde{\xi}_{0\kappa})]
\end{equation}
\begin{equation}
\begin{split}
	\Lambda_2 = & \left(\frac{\kappa_s + \frac{3}{2}}{\kappa_s-\frac{3}{2}}\right)\left(\frac{\kappa_s +\frac{1}{2}}{\kappa_s-\frac{3}{2}}\right)\xi_0^2\left(1-\frac{\xi_0^2}{3}\right)^2[1+\hat{\xi}_{0\kappa}Z_{\kappa+3}(\hat{\xi}_{0\kappa})] -\frac{1}{24}\left(\frac{\kappa_s - \frac{3}{2}}{\kappa_s-\frac{1}{2}}\right)+ \frac{1}{2}\left(\frac{\kappa_s + \frac{1}{2}}{\kappa_s-\frac{3}{2}}\right)\left(1- \frac{\xi_0^2}{3}\right)^2\\
	& -\left(\frac{\kappa_s + \frac{1}{2}}{\kappa_s-\frac{3}{2}}\right)(1-\xi_0^2)\left(1-\frac{\xi_0^2}{3}\right)[1+\tilde{\xi}_{0\kappa}Z_{\kappa+2}(\tilde{\xi}_{0\kappa})] + \frac{1}{6}\left(1-\frac{\xi_0^2}{2}\right) - \frac{\Psi_1(\kappa_s)}{2}[1 + \bar{\xi}_{0\kappa}Z_{\kappa+1}(\bar{\xi}_{0\kappa})]\,.
\end{split}
\end{equation}

Finally, for completion, for the core population described by a bi-Maxwellian distribution \eqref{eq_core_dist}, the susceptibilities $\chi_i$ are given by

\begin{equation}
	\chi_1(f_c) = -\frac{\omega_{pc}^2}{4\pi\omega^2} + \frac{1}{8\pi}\frac{\omega_{pc}^2}{\omega^2}\left(\frac{\alpha_{\perp}}{\alpha_{\parallel}}\right)^2\sum_{n=-1,1}\left[1+\phi_nZ(\zeta_n)\right],\\
\end{equation}

\begin{equation}
	\chi_2(f_c) = \frac{i}{8\pi}\frac{\omega_{pc}^2}{\omega^2}\left(\frac{\alpha_{\perp}}{\alpha_{\parallel}}\right)^2\sum_{n=-1,1}n\left[1+\phi_nZ(\zeta_n)\right],\\
\end{equation}
and
\begin{equation}
\chi_3(f_c) = \frac{1}{2\pi}\frac{\omega_{pc}^2}{k^2\ \alpha_{\parallel}^2}\left[1+\zeta_0Z(\zeta_0)\right].
\end{equation}

In all of the above expressions $\omega_{pj}^2 = 4\pi n_j q_j^2/m_j$ is the square of the plasma frequency of population $j$, and  $Z(\zeta)$ is the plasma dispersion function, given by:
\begin{equation}
Z_(\zeta) = \frac{1}{\sqrt{\pi}} \int_{-\infty}^{\infty}\frac{e^{-t^2}}{t-\zeta}\ dt\,.
\end{equation}
In addition, $Z_{\kappa}$ is the modified plasma dispersion function~\citep{HellbergMace2002,vinas2015,vinas2017linear,moya2020towards}
\begin{equation}
Z_{\kappa}(\xi) = \frac{\Gamma(\kappa)}{\sqrt{\pi \kappa}\ \Gamma(\kappa - 1/2)} \int_{-\infty}^{\infty}\frac{\left(1+\dfrac{t^2}{\kappa^2}\right)^{-\kappa}}{t-\xi}\ dt\,,
\end{equation}
that, for any real value of $\kappa$ such that $\kappa>1/2$, can be expressed in term of the Gauss Hypergeometric function $_2 F_1$:
\begin{equation}
Z_{\kappa}(\xi) = i\frac{\kappa-1/2}{\kappa^{3/2}}\,_{2}F_1\left[1,\, 2\kappa;
, \kappa+1;\, \frac{1}{2}\left(1-\frac{\xi}{i\kappa^{1/2}}\right)\right]\,.
\end{equation}
Finally, for $n=-1,0,1$ we have defined the following parameters: 
$$ \xi_n = \frac{\omega-n\Omega_s}{k\theta_{\parallel}},\ \ \ \ \varphi_n = \xi_n + \frac{n\Omega_s u_{\parallel}}{k\theta_{\perp}^2}$$
$$ \zeta_n = \frac{\omega-n\Omega_c - kU_c}{k\alpha_{\parallel}},\ \ \ \ \phi_n = \zeta_n + \frac{n\Omega_c \alpha_{\parallel}}{k\alpha_{\perp}^2}$$
$$\xi_{n\kappa} = \xi_n\sqrt{\frac{\kappa}{\kappa-3/2}},\ \ \ \ \bar{\xi}_{n\kappa} = \xi_n\sqrt{\frac{\kappa+1}{\kappa-3/2}},\ \ \ \ \tilde{\xi}_{n\kappa} = \xi_n\sqrt{\frac{\kappa+2}{\kappa-3/2}},\ \ \ \ \hat{\xi}_{n\kappa} = \xi_n\sqrt{\frac{\kappa+3}{\kappa-3/2}}\,,$$ 
$$\varphi_{n\kappa} = \varphi_n \sqrt{\frac{\kappa}{\kappa-3/2}}\,,$$

where $\Omega_j = q_j B_0/m_j c$ is the gyrofrequency.

\bibliography{arxiv}
\bibliographystyle{aasjournal}

\end{document}